\documentclass[twocolappendix]{emulateapj}
\usepackage{graphicx}
\usepackage{epstopdf}
\usepackage{amsmath}

\def\etal{\hbox{\em et al.}$\,$}

\def\Ha{\hbox{H$\alpha$}$\,$}
\def\Hb{\hbox{H$\beta$}$\,$}

\def\Msun{\rm{\hbox{M$_{\odot}$}}}           

\def\24m{\hbox{24~$\micron$}$\,$}
\def\22m{\hbox{22~$\micron$}$\,$}

\def\bt{\hbox{$B/T$}$\,$}
\def\r1{\hbox{$\rho_{1.0}$}$\,$}

\begin{document}

\slugcomment{Submitted to the Astrophysical Journal: 2016, Nov 16}

\title{The Star Formation Histories of Disk Galaxies: the Live, the Dead, and the Undead}

 \author{Augustus Oemler Jr\altaffilmark{1}, Louis E. Abramson\altaffilmark{2}. Michael D. Gladders\altaffilmark{3}, Alan Dressler\altaffilmark{1}, Bianca M.\ Poggianti\altaffilmark{4}, \& Benedetta Vulcani\altaffilmark{5}}

\altaffiltext{1}{The Observatories of the Carnegie Institution for Science, 813 Santa Barbara St., Pasadena, California 91101-1292}
\altaffiltext{2}{Department of Physics \& Astronomy, UCLA, 430 Portola Plaza, Los Angeles CA 90095-1547}
\altaffiltext{3}{Department of Astronomy \& Astrophysics, University of Chicago, Chicago, IL 60637} 
\altaffiltext{4}{INAF-Osservatorio Astronomico di Padova, vicolo dell'Osservatorio 5, I-35122 Padova, Italy}
\altaffiltext{5}{School of Physics, The University of Melbourne, VIC 3010, Australia}

\begin{abstract}
We reexamine the systematic properties of local galaxy populations, using published surveys of star formation, structure and gas content. After recalibrating star formation measures, we are able to reliably measure specific star formation rates well below that of the so--called ``main sequence'' of star formation vs mass.  We find an unexpectedly large population of galaxies with star formation rates intermediate between vigorously star--forming main sequence galaxies and passive galaxies, and with gas content disproportionately high for their star formation rates. Several lines of evidence suggest that these {\em quiescent} galaxies form a distinct population rather than a low star formation tail of the main sequence. We demonstrate that a tight main sequence, evolving with epoch as it is observed to do, is a natural outcome of most histories of star formation and has little astrophysical significance, but that the quiescent population requires additional astrophysics to explain its properties. Using a simple model for disk evolution based on the observed dependence of star formation on gas content in local galaxies, and assuming simple histories of cold gas inflow, we show that the evolution of galaxies away from the main sequence can be attributed to the depletion of gas due to star formation after a cutoff of gas inflow. The quiescent population is composed of galaxies in which the density of disk gas has fallen below a threshold for disk stability. The evolution of galaxies beyond the quiescent state to gas exhaustion and a complete end of star formation requires another process, probably wind--driven mass loss. The SSFR distribution of the quiescent and passive galaxies implies that the timescale of this process must be long, greater than a few Gyr, but less than a few tens of Gyrs. The environmental dependence of the three galaxy populations is consistent with recent numerical modeling which indicates that cold gas inflows into galaxies are truncated at earlier epochs in denser environments. 

\end{abstract}


\section{Introduction}

That the properties of many galaxies have evolved during recent epochs is a long established fact (Butcher \& Oemler 1978, BO78, Madau  1998, Cowie \etal  1996). That the end point of this evolution results in galaxy populations which vary over space has been known for even longer (Hubble 1936, Abell 1965, Dressler 1980, D80). Many processes may contribute to such evolution, of which the most fundamental are internal to the galaxy. Star formation builds structure, so structure will evolve as stars form. Star formation also depletes its own raw material so- unless provided with an unlimited supply of new gas (q.v. Larson, Tinsley, \& Caldwell 1980, LTC80)- star formation will necessarily drive its own decline and eventual extinction. However, the spatial variation in galaxy populations demands that external effects have significant influence on these processes. It is possible that this influence was exerted during the epoch of formation, setting up those internal properties of the protogalaxy which drove its subsequent evolution (e.g. D80). However, the fact that at least some of the environmental differentiation has occurred at recent epochs (BO78, Dressler \etal 1997, Allington-Smith \etal 1993, McGee \etal 2011 ) suggests- although it does not require- that much evolution is driven by the continuing interaction of galaxies with their surroundings. Many such processes have been suggested, including shutoff of the external gas supply (LTC80), stripping of gas by a hot external medium (Gunn \& Gott 1972), tidal encounters (Richstone \& Malmuth 1983, Byrd \& Valtonen 1990, Moore \etal 1996), and  mergers with other galaxies (Dressler \etal 1999, Struck 2006).
 
 Thus described, understanding galaxy evolution is a choice between galaxy properties set at birth and galaxy properties transformed by later interactions and, if the latter, between a number of distinct processes of transformation. Although the causes of the original galaxy properties may remain mysterious, their latter evolution seems a tractable problem. However, the success of the $\Lambda$ cold dark matter model, and rapid progress in high resolution modeling of galaxies in formation has shown that the simple choices outlined above are an  inadequate description of the actual histories of galaxies. Because the infall of baryons into dark matter halos and the hierarchical merging of these halos is an ongoing process, there are no clear distinctions between galaxy formation and galaxy evolution, or between galaxies and their surroundings. Recent numerical models of galaxy growth  have highlighted the involvement of  a number of physical processes, including infall of cold and hot gas, heating and possible expulsion of gas by stellar radiation, stellar mass loss, supernovae, shocks, and AGN's, and the transformation of gas content and distribution of stars by mergers. (The literature on this subject is now quite large; good recent summaries include Elahi \etal 2016, with and emphasis on early--type clustered galaxies, and Murante \etal 2014, emphasizing disk galaxies.) Although these models have had increasing success in reproducing some of the observed properties of galaxies, including mass functions, star formation histories, and even metallicities,it is still difficult to reproduce all of the properties of individual galaxies and galaxy populations. 
  
 A very different phenomenological approach to the same questions has become very popular during the last decade, inspired by the discovery of an apparent ``main sequence'' of galaxies in the star formation vs mass plane (Noeske \etal 2007).  Such a tight correlation between star formation and mass has suggested to some authors (e.g. Peng \etal 2010, Dutton \etal 2010, Sparre \etal 2015, Rodr{\'{\i}}guez-Puebla \etal 2016, Tacchella \etal 2016) that the myriad physical processes mentioned above  somehow conspire to produce {\em one} evolutionary track for all galaxies, that track being the main sequence. The majority of galaxies remain on that track even today, but a significant number have been ``quenched''; driven off the main sequence by one or more processes, including perhaps such classic mechanisms as stripping, starvation, etc. This idea has an obvious appeal; as is true in stellar astrophysics, the existence of a main sequence might point to a unique set of astrophysics which could bring order to the complicated mess which galaxy astrophysics appears to be. Whatever its appeal, it has yet to be demonstrated that it is correct. In a series of recent papers (Gladders \etal 2013, hereafter G13, Abramson \etal 2015, 2016 Dressler \etal 2016) we have argued for an alternative view. Motivated by the demonstration (Oemler \etal 2013) that many galaxies reached their star formation peaks at recent epochs, we have emphasized the {\em diversity} of the star formation histories of galaxies. In G13 this diversity was parameterized by log-normal star formation histories with two independent timescales, corresponding to the epoch of peak star formation, and the width of the peak. We have argued that such a description of galaxy histories conforms to the observations at least as well as the main--sequence evolution model, including explaining the existence of the main sequence itself.
 
 We are faced, then, with two sets of questions: how do the many astrophysical processes enumerated above result in the variation with time and environment of galaxy properties, and do these processes lead to a unique path of evolution or to many diverse paths? This is a daunting task, but the recent history should be the easiest aspect of galaxy evolution to understand, for several reasons.  With some oversimplification, the complex astrophysics of galaxy formation and evolution can be divided into two rather distinct areas of (1) extrinsic effects which govern the infall of gas into the regions in which stars are forming and (2) intrinsic effects which determine the fate of the gas once it arrives there.  Most of the difficulties in the correct modeling of galaxy formation lie in the intrinsic processes.  It has been clear since the earliest simulations of forming galaxies in a cold dark matter universe ( White \& Frenk 1991, Navarro \& Benz 1991)  that successful models of real galaxies must include massive amounts of thermal and/or mechanical feedback to disrupt infalling cold gas. Models with inadequate feedback inevitably turn too much of the initial baryons into stars, resulting in galaxies which form too fast, are too massive, and are too centrally concentrated. Getting the feedback right has been a difficult task, one that is not yet complete (see  Scannapieco 2012,  and Elahi \etal 2016,  for comparisons of various recent attempts). This difficulty is usually attributed to the inevitably incomplete resolution of small--scale astrophysical processes- such as the interaction of stellar winds and radiation with the very inhomogeneous interstellar medium- which are central to the feedback processes.  The extrinsic regime should be less troublesome, since it is dominated by the dynamics of dark matter and of baryons in a much less non--linear regime than that which occurs within galaxies themselves.  At low and moderate redshifts most galaxies seem to be in a quasi-equilibrium state, with ordered internal structure and dynamics, and moderate to low levels of star formation. However incomplete our understanding may be of the detailed astrophysics of the intrinsic processes of feedback and star formation, the results of these processes can be observed directly in nearby galaxies in such forms as the Schmidt-Kennicutt law (Schmidt 1959, Kennicutt 1989), which has been shown to govern star formation in a wide variety of galaxies at a range of redshifts (e.g. Daddi \etal 2010, Bouche \etal 2007), and which recent numerical models (e.g. Hopkins \etal 2014) are now able to reproduce  with some accuracy. We may be left, then, with only the extrinsic effects as possible variables to explain the recent histories of galaxies.
  
There is, of course, a very large literature on the systematic properties of nearby galaxy populations. It is well established that there are strong trends present in these populations, including trends of star formation rates with mass, epoch, structure, and environment, of metallicity with mass  and galaxy type, and of structure with mass and environment. Also, much work has gone into attempts to delineate the ``main sequence'' and its evolution with time (Speagle \etal 2014 present a compilation of many of these).  Despite all of this work, the subject of local galaxy populations and their implications for galaxy evolution is far from exhausted. In this and subsequent papers we use multiple published surveys of the local population of galaxies to reexamine the systematic behavior of their main observable properties: mass, star formation rate, disk scale, bulge fraction and gas content. Our primary goal in this first paper is to understand the present distribution and recent evolution of gas and star formation in disk galaxies. A subsequent paper will examine the issue of bulges and disks. After a careful recalibration of star formation rates, we will show that we can reliably measure star formation rates well below the main sequence, and thus follow galaxies as they evolve from the main sequence towards the passive state. We will show that the well--established bimodality in galaxy colors hides an important intermediate population of not-quite-dead galaxies with significant gas but low star formation rates. Delineating the systematic properties of the star--forming, dead, and not-quite-dead galaxies will allow us to put useful new constraints of the processes governing recent galaxy evolution.
 
The layout of this paper is as follows. In Section 2 we survey the Mass--star formation plane, reaching the tentative conclusion that it can be subdivided into three distinct groups: the main sequence, non-star forming objects, and the intermediate group with low but non-zero star formation rates, which comprises about a third of the galaxy population. In Section 3 we assemble facts about these three populations, deferring most analysis until Section 4, in which we attempt to understand these properties in light of various theories of galaxy evolution. Section 5 summarizes the results. In the Appendix we will critically examine the star formation rates from Brinchmann \etal (2004), and recalibrate them using the best available data . This section is critical for establishing the reliability of our detection of low star formation rates, but can be skimmed by those mostly interested in the results.

\section{Mapping the Star Formation Rate-Mass Distribution}

\subsection{Deconstructing the Star Formation Main Sequence}

Brinchmann \etal(2004, B04) have derived star formation rates from imaging and spectroscopic observations of 149,660 galaxies from DR2 of the Sloan Digital Sky Survey (York \etal 2000) which have spectroscopic observations,  r magnitudes in the range $14.5 < r < 17.77$ and redshifts between 0.005 and 0.22.They also incorporate mass determinations from Kauffmann \etal (2003).  In 2007 they recalibrated their method, and applied it to the DR7 release. These new data are not published but  are available on the web.\footnote{http://wwwmpa.mpa-garching.mpg.de/SDSS/DR7/}. We will use this latter data set, which we denote as B04-DR7, in all of the following analysis. The mass values we shall use as given; all determinations of galaxy stellar masses are derived from SED data using stellar population models and we have no reason to think that we can improve of those of Kauffmann \etal. Any significant errors are probably systematic to the population model and will only shift the mass scale, which should not affect our analysis. There are, however, independent methods of determining star formation rates, and in Appendix A we use what we believe are the best of these to improve on the B04-DR7 SFR's. The main changes are to the star formation rates of galaxies with weak or undetected emission lines. Unless otherwise noted, all quoted SFR values are based on our recalibration.

 First, though, we begin with the original B04 data. A particularly striking presentation of  the star forming main sequence to be found in the B04 data is presented in Figure 1 of Peng \etal (2010, P10). We repeat this presentation of the B04 data in Figure 1. The density of objects is normalized so that the integrated density per interval of log  mass is constant, which permits a better appreciation of the behavior of star formation as a function of mass. This is a quite striking correlation, with a  scatter $\sigma(log(SFR)) \sim 0.29$.  If, as has often been assumed (but we show in Appendix A to be incorrect), the typical errors in B04 star formation rates were as large as 0.2 dex to 0.3 dex, Figure 1 would seem to imply that the true scatter in the star forming main sequence is quite small.

\begin{figure}[tbph!]
\figurenum{1}
\plotone{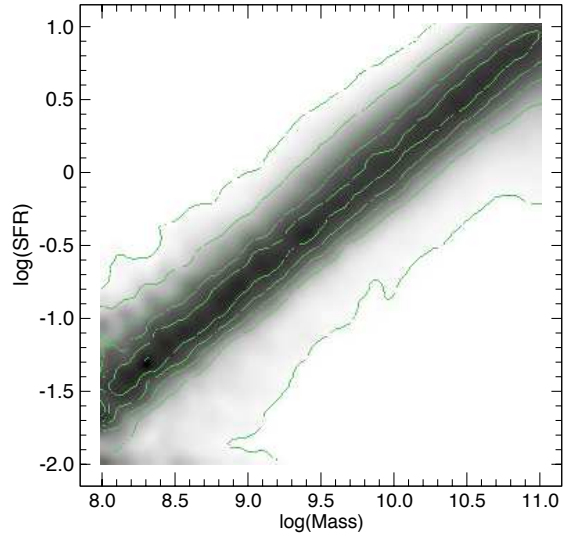}
\caption{Density of galaxies in the log(SFR)-log(Mass) plane, for the sample of B04 galaxies analyzed by P10. The contours represent equal normalized density intervals.}
\end{figure}

Unfortunately, this plot does not stand up to close scrutiny. The first thing to note is that plotting observed SFR versus stellar mass gives a quite misleading impression of the correlation between these quantities, because the stellar mass is simply the time integral of the star formation rate. More precisely, 
\begin{equation}
{M_{tot}} = \int\limits_{0}^{{t_0}} {SFR(t)f_ldt}  = f_lt_0\langle SFR\rangle 
\end{equation}
where $f_l$  is the fraction of the mass that remains locked up in stars, rather than returning to the ISM during stellar evolution. Unless $\langle SFR  \rangle$ is completely uncorrelated with  $SFR(t_0)$-- which would imply a completely random history of star formation-- SFR will {\em always} be correlated with mass, whatever the underlying astrophysics or history of star formation. A less misleading presentation of the B04 data can be obtained by plotting {\em specific} star formation rate, $SSFR = SFR/M_{tot}$ versus mass. we can write SSFR as

\begin{equation}
SSFR(t) = \frac{1}{{f_lt}}\frac{{SFR(t)}}{{\left\langle {SFR} \right\rangle }}
\end{equation}

Thus, SSFR is an (incomplete) measure of the star formation history of a galaxy, in that, while many histories can lead to the same value of SSFR(t), different values of SSFR(t) require different histories.

Also, we must improve on the data set used in Figure 1. Firstly, we replace the B04 star formation rates with those obtained in the Appendix, which are more accurate, particularly for low star formation rates.   Secondly, P10 cull their galaxy set in two problematic ways: by excluding galaxies without strong emission lines or with some sign of an AGN component, and by excluding galaxies without sufficiently blue $U-B$ colors. All of these exclusions will narrow the SSFR range. The result of these corrections is presented in Figure 2 where we now plot the SSFR-log(Mass) distribution for all objects with detected star formation (i.e. $logSSFR>-12.2$). (We now plot individual points rather than a normalized distribution.) 

\begin{figure}[tbph!]
\figurenum{2}
\plotone{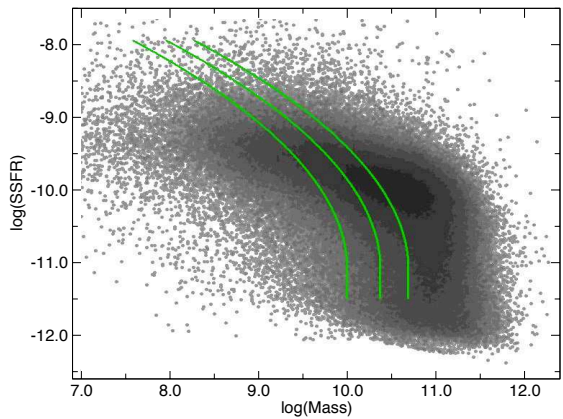}
\caption{The SSFR vs Mass relation for the entire B04-DR7 sample, using the SSFR values derived in Appendix A. Superimposed are the median completeness limits at, from left to right, the first through third quartiles of the redshift distribution of the galaxies. }
\end{figure}

This distribution is obviously much broader than that in Figure 1, but one source of bias remains. The analyses of B04 and P10, as well as those presented so far in this paper, have included all sample galaxies (of particular spectral properties) at every redshift. The SDSS spectroscopic sample is complete to $r=17.8$. To see what this means for mass completeness we must consider the mass-to-light ratios of the galaxies. In Figure 3 we plot the distribution of r-band mass to light ratios of the B04-DR7  galaxies versus their SSFR, and  fit a polynomial to the median  value of M/L(r) vs log(SSFR), over the SSFR range $-11.5 \le log(SSFR) \le -9.2$ where most of our objects lie.  Superimposed on the SSFR-Mass distribution in Figure 2 are the mass limits derived using that polynomial fit  at the first through third quartiles of the redshift distribution of the sample, namely  z = 0.054, 0.083, 0.119.  The sample is more than 50\% incomplete to the left of each line. Clearly, the data as usually presented are seriously incomplete. Furthermore, because of the slope of the completeness lines, galaxies with lower SSFR's are systematically excluded since they are less luminous at a given mass than those with high SSFR's. Note that this effect operates at every mass, since the higher mass galaxies are drawn preferentially from higher redshifts.

\begin{figure}[tbph!]
\figurenum{3}
\plotone{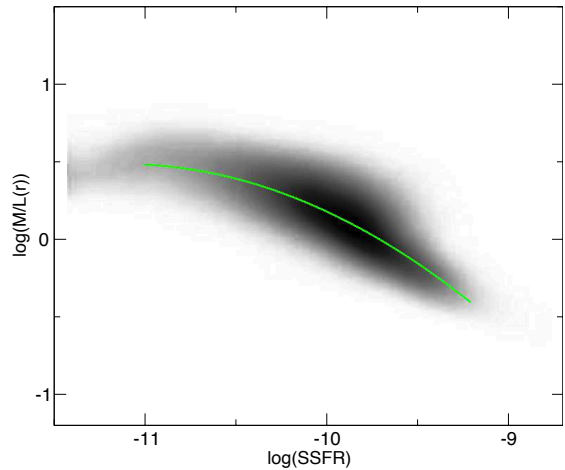}
\caption{r--band mass to light rations of the B04 sample, vs SSFR. The green line is the adapted median relation. }
\end{figure}

\begin{figure}[tbph!]
\figurenum{4}
\plotone{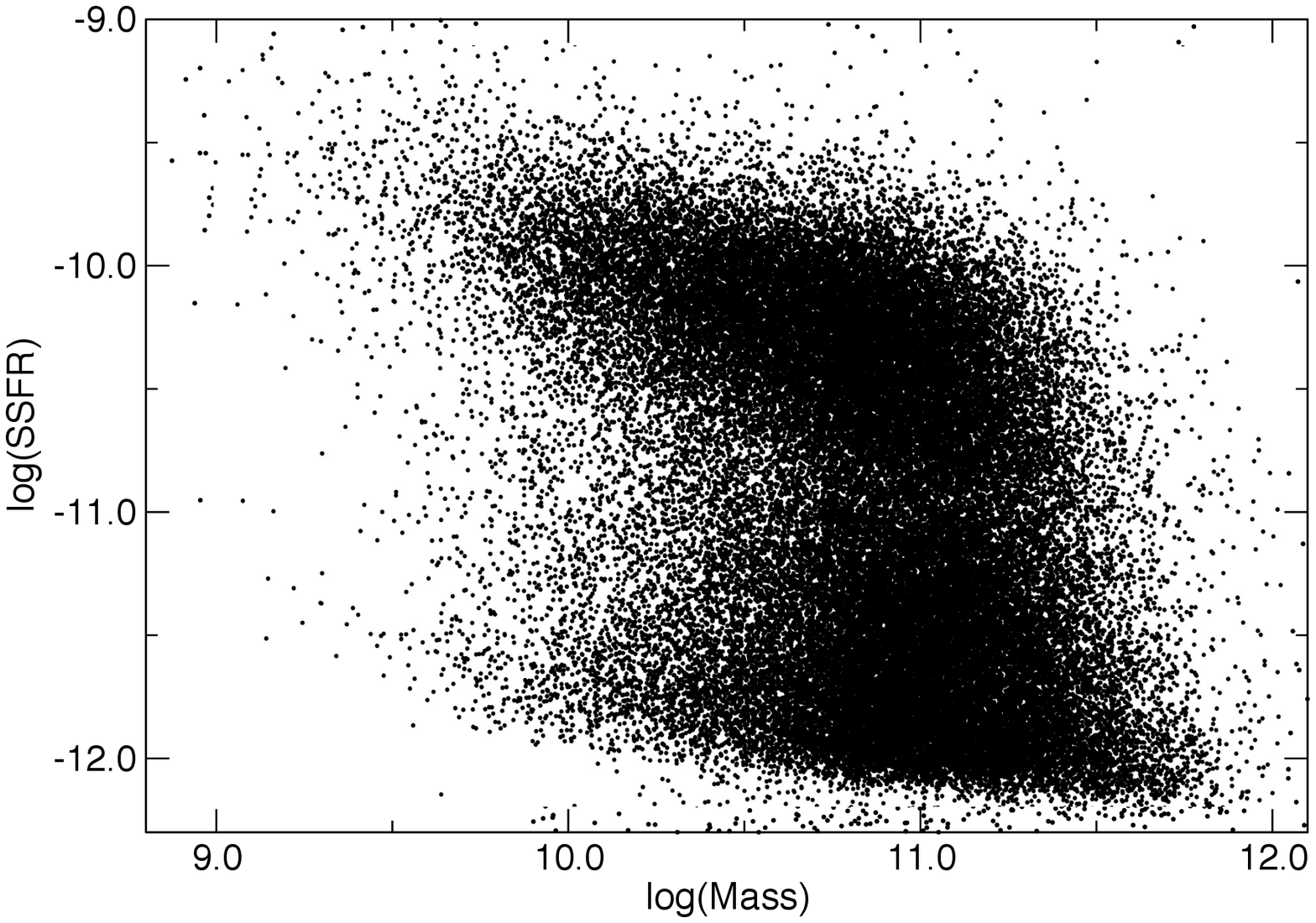}
\caption{SSFR versus mass for the maximally complete sample.}
\end{figure}

Given the results in Figures 2 and 3, we can construct the largest complete sample of B04-DR7, which we denote the {\em maximally complete sample}, by, at every mass, including only galaxies out to a redshift such that all galaxies with $log(M/L(r)) \le 0.8$, the maximum found in the sample, will have $r\le17.8$, and therefore will be included in the complete SDSS spectroscopic sample. In order to minimize the mismatch between galaxy size and SDSS fiber size, we also limit the sample to objects with $z \ge 0.01$. Note that this sample, although unbiased in star formation properties at each mass,  is strongly biased towards high mass, since the sample of higher mass objects is complete over larger volumes. 

The results for this sample are presented in Figure 4. The beautifully clean star--forming main sequence of Figure 1 has almost disappeared. Although some sort of sequence is apparent at low mass, it seems to be entirely obscured at high masses by a broad swath of lower SFR objects. In all following analyses, the sample will be limited to the maximally complete sample unless explicitly stated otherwise. 

\subsection{Repopulating The Mass--Star Formation Plane}

Figure 4 shows that the SSFR--Mass distribution is very broad, with a distinct narrow main sequence only apparent at low Masses ($log(Mass) < 9.5$). However, closer examination shows that this broad distribution has significant structure. Figure 5 presents the distribution of $SSFR$  in three intervals of log(Mass): 9.3--9.7, 9.8--10.2, and 10.3-10.7. Three peaks are apparent in all of these: one for those without detectable star formation- which we arbitarily place at $log(SSFR) = -12.8$ , one at $log(SSFR) \sim 11$, and one at $log(SSFR) \sim 10$.  The middle peak is broad, and does not shift with mass, but the upper peak is narrow, and moves to lower values of log(SSFR) as mass increases. It is reasonable to still call the upper peak the ``main sequence''. Using all objects with $log(SSFR) > log(SSFR)_{peak} -0.5$ , we obtain the dispersion of the main sequence - $\sigma(log(SSFR)) = 0.30$, independent of Mass, as well as its slope $d(log(SSFR_{peak}) /d(logMass) = -0.43$. It should be noted that the width of the main sequence, though narrow, is real, since the scatter in our best SSFR measure, established in Appendix A, is substantially less than 0.3dex at the main sequence.

\begin{figure}[tbph!]
\figurenum{5}
\plotone{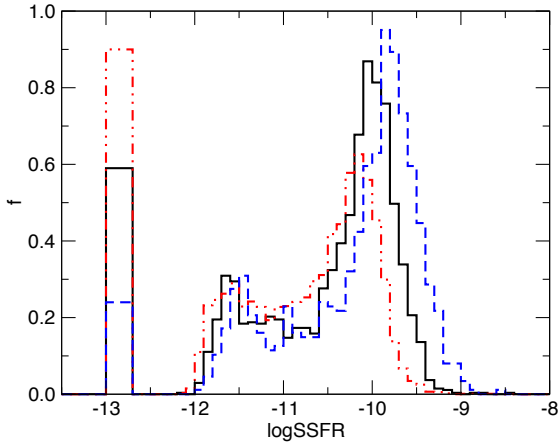}
\caption{Distribution of log(SSFR) for galaxies in the maximally complete sample, in three narrow intervals of log(Mass), centered at 9.5 (blue), 10.0 (black), and 10.5(red). }
\end{figure}

Let us assume for the moment  that the main sequence is a distinct component of the SSFR--Mass distribution, separate from the intermediate SSFR group, with the shape  described above.  If we also assume that the main sequence dominates the distribution at values of log(SSFR) higher than the main sequence peak,  and that the distribution is symmetrical about the peak, we can use counts of objects with $SSFR > SSFR_{peak}$ to obtain its mass function, and using that plus the slope and spread, we can  generate the entire main sequence  and statistically remove it from the SSFR--Mass distribution of all galaxies. The results of this decomposition-ignoring the dead galaxies with no detectable star formation- are presented in Figure 6. The most striking implication of this decomposition is that the main sequence, rather than dominating the SSFR-Mass plane, as implied by, for example, Figure 1, only contains about half of the star--forming galaxies. The other half belong in a broad swath stretching from below the main sequence down the the level at which star formation becomes undetectable.

\begin{figure}[tbph!]
\figurenum{6}
\plotone{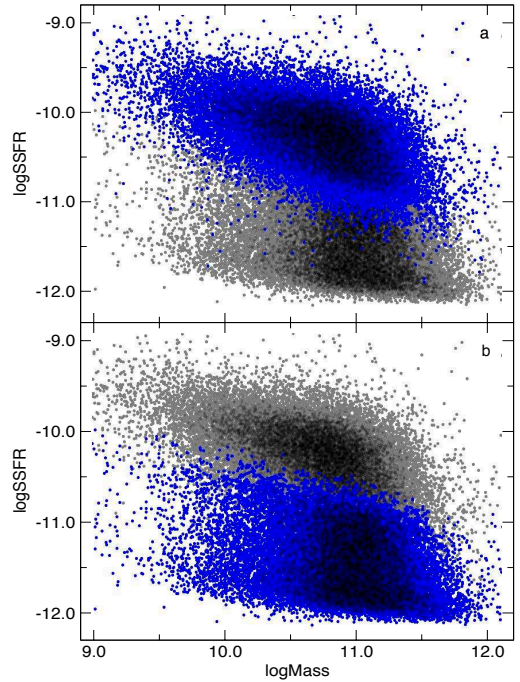}
\caption{Decomposition of the maximally complete sample of star--forming galaxies into main sequence galaxies and others. In (a) the main sequence objects are shown in blue and overlay the others, in (b) the order is reversed. }
\end{figure}

The existence of a population of galaxies with star formation rates significantly lower than the main sequence is not a new discovery, but it has not received much attention. It is absent from color distributions, which only show the main sequence and no--star--formation peaks and an almost empty ``green valley'', because galaxy colors are insensitive to SSFR values below about  $log(SSFR) \sim -11$. (q.v. Dressler \& Abramson 2015) Such a population is apparent in Figure 24 of B04, as well as in Figure 22 of Salim \etal (2007).  Schiminovich \etal (2010) discuss it in relation to star formation efficiency (to which we will turn in Section 3.3), and Tojeiro \etal (2013) in relation to star formation histories. In most cases its importance has been underestimated because of the sample bias described by Figure 2, as well as by the difficulty in calibrating very low star formation rates.  SDSS images  of a random subset of these low SSFR galaxies are presented in Figure 7. Most appear to be rather normal early--type galaxies. A set of 51 galaxies in the low--star formation group which are members of the Coma Cluster and were classified by D80 contains only 4 that were not typed as normal E's or S0's, including 1 Ep, 2 S0/a's, and 1 Sb.

\begin{figure*}[tbph!]
\figurenum{7}
\plotone{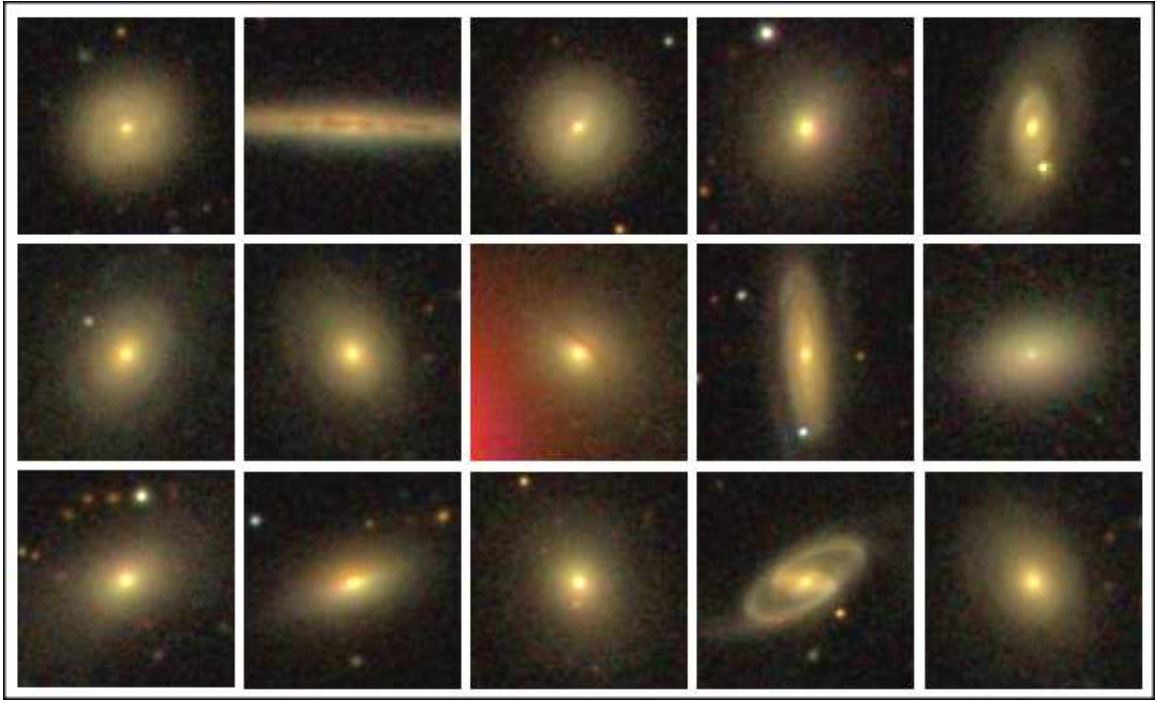}
\caption{SDSS images of a representative sample of brighter galaxies with $log(SSFR) \sim -11.5$. Each image is $48 \arcsec$ across.}
\end{figure*}

The identification of the low SSFR subset of galaxies as a distinct component is only a convenient working hypothesis, although the distributions in Figure 5 give some credibility to the idea. Also, the distinction between these objects and those with no detected star formation is presumably an artifact of our inability to measure specific star formation rates below $log(SSFR) = -12$. Nevertheless, for the moment we will divide the galaxy population into three distinct groups: those with vigorous star formation which we will continue to call the {\em main sequence}, the intermediate group, which we will denote as {\em quiescent}, and those with no detected star fomation, which we will call {\em passive}. For the following analysis, we will label galaxies as "main sequence" or as "quiescent" according to their assignment in the statistical decomposition presented in Figure 6. In the overlap region this is, of course, an arbitrary assignment, but no more arbitrary that doing the cut at some specified value of SSFR. All objects below the detection limit of $log(SSFR) = -12.2$ are classified as "passive." In coming sections, we  will explore how the properties of these populations are related to their gas content, environment, and structure.

\section{Properties of the Three  Populations of Galaxies}

\subsection{Environmental Dependence}

 \begin{figure}[tbph!]
\figurenum{8}
\plotone{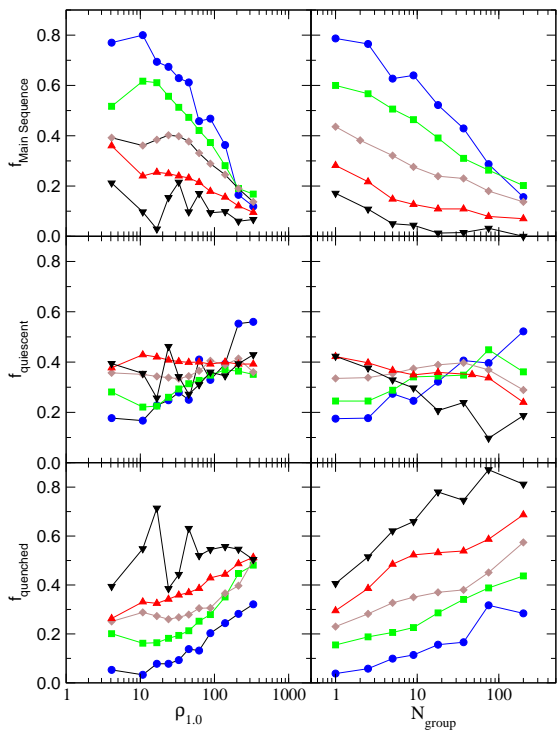}
\caption{Variation of fractions of the three galaxy types with environment,  in 5 ranges of log(Mass); blue circles: 9.5 -- 10.0, green squares: 10.0 -- 10.5, brown diamonds: 10.5 -- 11.0, red up trianges: 11.0 -- 11.5, black down triangles: 11.5 -- 12.0. Left column- galaxy populations vs galaxy density within 1 Mpc; right column- populations vs group membership. Top row- fraction of Main Sequence objects, middle row- fraction of quiescent objects, bottom row- fraction of passive objects.}
\end{figure}

In Figure 8 we present the incidence of our three star formation classes as a function of environment, in five mass ranges, using two measures of environment. We take these measures from Tempel \etal (2012) who, as part of a clustering survey of SDSS DR8 galaxies, have determined group membership and normalized galaxy densities within 1, 2, 4, 8, and 16 Mpc of a sample of SDSS galaxies which includes about 89\% of our maximally complete sample; we will use their group membership, as well as density within 1 Mpc, \r1. Errors for points in the left column are typically 0.01, those in the right column range from 0.005 for small $N_{group}$ to 0.02 for $N_{group}>100$. It is clear that star formation is sensitive to environment at every mass and in every environment, though high--mass galaxies show a lower sensitivity than do low--mass ones.  Conversely, in every environment high--mass galaxies are more likely to be quiescent or passive than are low mass objects. At first glance, the population of quiescent galaxies appears to be quite stable, with only the ratio of star--forming to passive galaxies varying with environment, but this appearance is misleading. In Figure 9 we plot, for every data point in Figure 8, the fraction  of main sequence  galaxies versus the ratio of quiescent to passive galaxies. The result shows that there is a very strong correlation between these two quantities at every mass and local density. The lower the fraction of main sequence galaxies, the lower the fraction of galaxies below the main sequence with some star formation. 

\begin{figure}[tbph!]
\figurenum{9}
\plotone{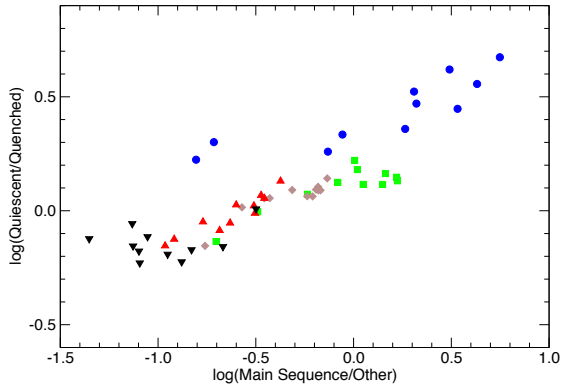}
\caption{Fraction of main sequence galaxies versus the ratio of quiescent to passive galaxies, for each data point in Figure 8. Points are labeled by mass range as in Figure 8. }
\end{figure}

\subsection{Gas Content}

Two HI surveys cover large portions of the volume of the B04-DR7 sample. The ALFALFA survey (Giovanelli \etal 2005, Haynes \etal 2011) used the Arecibo Telescope for a blind survey of a significant fraction of the North Galactic Cap over the velocity interval $1600 km s^{-1} \le cz \le 18000 km s^{-1}$. The survey volume included about $3 \times {10^4}$ B04-DR7 galaxies ($1\times 10^4$ in the maximally complete sample), of which about $9\times10^3$ were detected ($3\times10^3$ in the maximally complete sample). The sensitivity of the survey varies with linewidth, but a typical limit for 50\% completeness is $log(M_{HI}) = 12.63+2log(z)$
, or $5\times10^9\Msun$ at the median redshift of the survey. The GASS survey (Catinella \etal 2012) is a pointed survey of about $10^3$ galaxies within the ALFALFA survey volume which were {\em not} detected either by ALFALFA or by Springob (2005). It includes only objects in the mass range $10.0 \le log(Mass) \le 11.2$, and the redshift range $0.025 \le z \le 0.05$. It is about a factor of 5 deeper than ALFALFA at the same redshift.

Figure 10 summarizes the HI content of the B04-DR7 galaxies derived from these two surveys. Figure 10a plots the median gas to stellar mass ratio as a function of log(SSFR), in intervals of galaxy mass, and Figure 10b plots the $1\sigma$ dispersion, calculated from the difference between the 50th and 84th percentiles of the distribution. Above $log(SSFR) = -11.0$ the ALFALFA survey detected the majority of the galaxies over some subset of the survey volume, allowing a determination of both quantities.  Since the survey is approximately flux--limited, the detection rate is a function of $M_{gas}z^{-2} = M_{stars} (M_{gas}/M_{stars})z^{-2}$. The redshift limit of each $M_{stars}$  and SSFR subset is chosen to sample as large as possible a fraction of the ALFALFA volume, while still reaching deep enough into the $M_{gas}/M_{stars}$ distribution to allow determination of the median value. Since the detection threshold is rather broad, there will be some detections below the completeness limit and some non--detections above the limit. We handle non--detections in two ways (1) by assigning them a value of $M_{gas}/M_{stars}$ equal to the 50\% completeness limit or by assigning them values of zero. Where these differ, both points are plotted. When the majority of the objects are undetected, the latter value goes to zero, as with the leftmost black point in Figure 10a. The ALFALFA results are presented in Figure 10 as colored points, divided into the same mass intervals as in Figure 8. Errors in the ALFALFA sample range from 0.02 to 0.04 for median values of $log(M_{gas}/M_{stars})$, and from 0.015 to 0.035 for $\sigma$.

\begin{figure}[tbph!]
\figurenum{10}
\plotone{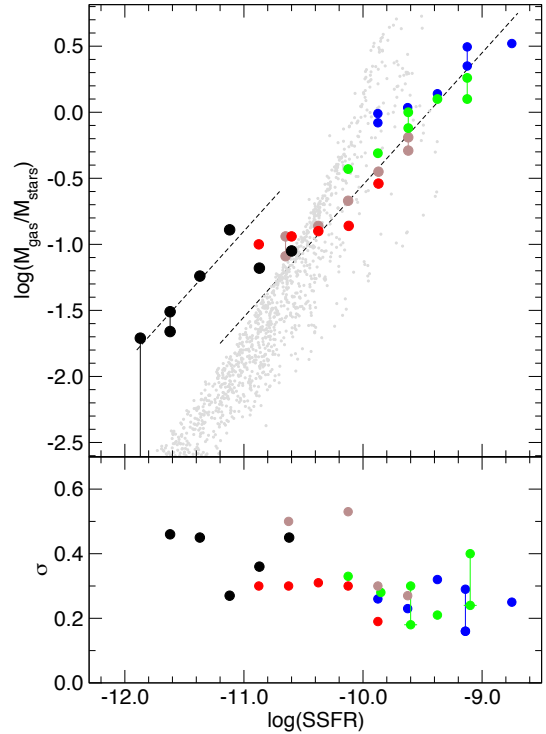}
\caption{HI gas content per unit mass versus SSFR. a- median gas content, b- $1\sigma$ dispersion about the median. black circles- GASS+ALFALFA sample, colored circles- ALFALFA sample in 4 ranges of log(Mass): blue-  9.0--9.5, green-  9.5--10.0, brown- 10.0--10.5, red- 10.5--11.0.   upper dashed line- $\tau_{gas} = 21 \times 10^9$years, lower dashed line-  $\tau_{gas} = 4.7 \times 10^9$years, grey points- log--normal model galaxies from G13.}
\end{figure}

Below $log(SSFR) = -11.0$ the ALFALFA sample in insufficiently deep and we turn to the GASS survey. Here things become slightly complicated. The GASS sample is a random subset of objects within the ALFALFA survey  volume with redshifts $0.025 \le z <0.05$ and masses $10.0 \le log(Mass) < 11.2$, {\em but} excluding objects that were already detected by ALFALFA. Thus, the GASS sample by itself is biased against high--HI content objects. To correct for this bias, we must add back those objects which would have been included if not detected by ALFALFA. Since the survey does not provide information on the number of non--included objects,  we must estimate it from the ALFALFA sample itself, by the following procedure. The GASS sample objects are (almost) all contained within our B04 Maximally Complete Sample, which we abbreviate MAXSAMP. Let $S_A$ be the subset of ALFALFA objects in MAXSAMP and within the GASS mass and z ranges. Let $S_G$ be the subset of GASS objects within MAXSAMP. Since the ALFALFA detection probability within  ASAMP will depend of galaxy mass, we divide the sample into mass intervals. Then, it is straightforward to show that, in each interval, if $\tilde A$ is the number of {\em undetected} $S_A$ objects, and $G$ is the total number of $S_G$ objects, we can combine the GASS objects with the ALFALFA detected objects by  giving each $S_A$ detection a weight of $G/\tilde A$. The GASS sample is too small to subdivide by mass, so we include all objects. We deal with GASS nondetections in the same manner as with the ALFALFA sample. The results are presented in Figure 10 as black points.  Typical errors  in the GASS sample are about 0.08 for the median values of $log(M_{gas}/M_{stars})$ and 0.06 for values of $\sigma$.

 The diagonal dashed lines are lines of constant gas--to--star formation rate, which is, to within a factor of $f_l$,  the inverse of gas consumption timescale, $\tau_{gas}$. The upper line corresponds to $\tau_{gas} = 21 Gyr$, and the lower line to $\tau_{gas} = 4.7 Gyr$. Several points seem clear. (1) At a given SSFR, the scatter in gas to stellar mass ratio is rather small, no more than a factor of 2. (2) gas consumption timescales seem to be independent of galaxy mass. (3) The trend of gas content with SSFR is not inconsistent with a simple picture in which {\em} the median gas consumption timescale of main sequence galaxies is 4.7 Gyr, independent of SSFR, and the median gas consumption timescale of quiescent galaxies is 21 Gyr, independent of SSFR.
 
  Another view of the gas consumption issues is illustrated by the gray points. In Gladders G13 we constructed models for the star formation histories of galaxies with log--normal shapes of the form\begin{equation}
 SFR(t) = {S_0}\frac{1}{{t\tau\sqrt {2\pi } }}{e^{ - {{(\ln t - \ln ({t_0}))}^2}/2\tau^2)}}
 \end{equation} fitting the distribution of shape parameters using observations of low and high redshift galaxy populations as constraints. If galaxies were closed boxes, with gas neither entering nor leaving, and if all gas was eventually turned into stars, then a galaxy's gas content at any time would be just that needed to form the mass of stars it will form in its future. Since we know the future evolution of each G13 galaxy, we can determine the gas--SSFR distribution of those galaxies. This is presented as grey points in Figure 10.  The important point is that the median of the grey points lies near or above the median gas content of main sequence galaxies, but well below that of the quiescent objects.  Thus, the median gas content is more than sufficient to fuel the future star formation of  all of the quiescent galaxies, but not all main sequence objects. These models represent only one possible description of galaxy formation histories, though they do fit well all of the extant constraints. However, it is likely that most alternate models, or at least those in which star formation decreases steadily from a peak at early times, will have a qualitatively similar distribution. Therefore, it appears that many (though far from all) main sequence galaxies will require a continuing inflow of gas to maintain their star formation,  but that quiescent galaxies do not. We will consider the implications of this fact in Section 4.

\subsection{Galaxy Structure}

We close this section with a brief look at galaxy bulges and disks. We shall examine this in much greater detail in a subsequent paper, but one important fact will be useful in the following discussion. It has long been known (Hubble 1935, Baade 1951) , that the masses, stellar populations, and structures of galaxies are correlated: high mass, bulge--domination, and low star formation rates tend to be associated with each other, as are low mass, disk--domination, and high SFR. Mendel \etal (2014) have determined the masses of disks and bulges for a large fraction of the B04-DR7 set of galaxies. In Figure 11 we compare the bulge mass fractions (which we will designate \bt) of the three galactic populations with masses in the range $10.0 \le log(Mass) < 11.0$. To avoid the ambiguous overlap region near $log(SSFR) \sim -10.75$ we only take main sequence galaxies with $log(SSFR>-10.5)$ and quiescent galaxies with $log(SSFR) < -11.0$. The import of this figure is clear: quiescent and passive galaxies are structurally almost identical and very different from the star--forming population. While star--forming galaxies have an almost uniform distribution of bulge mass fraction, {\em there are almost no disk--dominated quiescent or passive galaxies}, those with bulge mass fractions less than 0.2. Conversely, there are very few bulge--dominated ($B/T>0.9$) star--forming galaxies. Much but not all of this is old news: it is a commonplace that there are very few star--forming ellipticals, and the lack of disk--dominated S0's was first noted by  Sandage \& Visvanathan (1978), who argued from this that S0's could not be stripped spirals (more on this later.)  

\begin{figure}[tbph!]
\figurenum{11}
\plotone{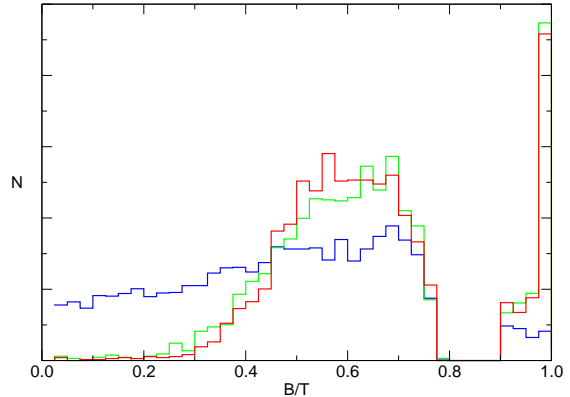}
\caption{Distribution of bulge mass fraction for star--forming- blue, quiescent- green, and passive-red galaxes, in the mass range $10.0 \le log(Mass) < 11.0$. }
\end{figure}

\section{Discussion}

\subsection{The Physical Significance of the Main Sequence}

Multiple lines of evidence  seem to suggest that main sequence and quiescent galaxies form distinct populations. The evidence includes the SSFR distribution at fixed mass (Figure 5), the variation in the incidence of the populations with environment (Figures 8),  the SSFR dependence of gas content (Figure 10), and the almost total absence of pure disk galaxies among the quiescent and passive populations (Figure 11). That the main sequence is a physically distinct population has been a common theme of many recent papers, many of which have tried to identify the astrophysical processes responsible for the tight SFR vs mass dependence. The above list seems to support such a view, but the existence of a well--defined population does not necessarily imply a physical process driving the SFR--mass correlation. As was pointed out in Section 2, {\em any} star formation histories that are not completely random will produce a correlation of SFR with mass. As Equation 2 makes clear, the peak of the resulting SSFR distribution will be determined only by the age of galaxies ($t _0$) and the mean value of $SFR(t)/\left\langle {SFR} \right\rangle $. If the majority of  galaxies are of comparable age then only the range of the shapes of the star formation histories will produce a scatter in SSFR.

Two very general considerations require that the SSFR distribution fall rather rapidly away from the peak. Equation 2 constrains the high side. If we assume that very few galaxies have rising star formation rates at the present epoch, Equation 2 requires that the SSFR distribution fall fast at high values of SSFR, nearing zero at $SSFR(t) = 1/ft_0$. Similar considerations constrain the low side. At early times- shorter than the timescale over which SFR varies- $SFR(t) \approx \left\langle {SFR} \right\rangle $ for all galaxies. Therefore, if star formation rates are now falling, and if $d^2SFR/d t^2$ is zero or negative, the galaxies with lowest values of SSFR today will be those in which SFR has fallen the fastest.  If lower SSFR galaxies evolve faster, then galaxies accelerate away from the "main sequence", traversing each interval of SSFR more quickly, resulting in a decreasing density of objects at lower SSFR. Note that these effects depends not at all on any particular assumptions about the physical processes driving the star formation history. 

 As an illustration of this effect, we construct a random population of galaxies based on  the log--normal star formation histories, of the form described by Equation 3, developed in G13.  In that paper, the distribution of the parameters $t_0$ and $\tau$ were tuned to match observations of the history of the star formation density as well as of the SSFR distribution.  Here we take a completely random set of values over the range $1.5Gyr \le t_0 < 20Gyr$, $0.1 \le \tau < 1.1$, considerably broader than what was needed to fit the observed SSFR distributions. We evolve the histories until the present age of the universe and calculate the present epoch SSFR of  each objects. Values of $log(SSFR) <-12$ are assumed to be undetectable and are set to the arbitrary value of -13. The results are presented  in Figure 10; from the 16th and 84th percentiles of the distribution of detected objects we calculate $\sigma(log(SSFR) = 0.36$, only slightly broader than we have determined for the real ``main sequence'' in Section 2. Thus, a completely random set of star formation histories produces a SSFR distribution almost identical to that observed. The point is that the existence of a ``main sequence'' signifies little about the history of the star formation in a galaxy population. We have chosen a log normal form for illustration, but virtually any set of star formation histories satisfying the two constraints that (1) $d^2SFR/d t^2$ is zero or negative and (2) star formation histories are smooth rather than having large random excursions (which could temporarily push SSFR to arbitrarily high or low values) will produce a similar distribution. From Equation 2, we expect the peak to be located somewhat lower than $1/t$, which means that it will rise with increasing redshift, as it is observed to do. Therefore, the attempt to construct physical models to explain the properties of the main sequence is misguided. Although it is possible that some detailed astrophysical considerations could duplicate the behavior which we have seen to occur naturally, simplicity demands that we accept the straightforward explanation described above, until some new information forces us to look for elsewhere.
 
\begin{figure}[tbph!]
\figurenum{12}
\plotone{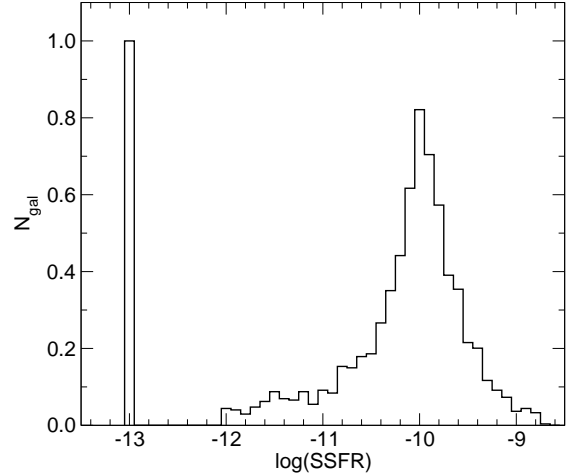}
\caption{The SSFR distribution of galaxies at the present epoch, from a simulation of log normal star formation histories in which the two parameters of the log normal are distributed at random over the range $1.5 \le t_0 < 20$Gyr, $0.1 \le \tau < 1.1$.}
\end{figure}
 
 \subsection{The Physical Significance of the Quiescent Galaxies}
 
 If the main sequence is a natural consequence of the age of galaxies and a smooth evolution of their star formation rates, the same is not true of the quiescent population. Figure 12, and the preceding argument, both show that there should be a rapidly falling tail of the SSFR distribution to the low side of the main sequence peak, rather than the large population of galaxies at low SSFR values that are observed. Although the existence of a population of intermediate SSFR galaxies is not news, many of the properties of this population are new, and are quite puzzling, so we should first address the possibility that our low SSFR galaxies are spurious, and the result of errors in the star formation rates. We demonstrated in Appendix A, by a comparison with star formation rates derived from (mostly independent) UV+IR measures,  that our SFR measure is quite accurate, but the possibility remains that {\em all} of the star formation measures are systematically wrong. 
 
 It is easy to rule out the possibility that the SSFR values of the quiescent objects are spuriously low, i.e. that they are misclassified main sequence objects. Neither their appearance (Figure 7), dependence on mass and environment (Figure 8), gas content (Figure 10), or structure (Figure 11) look anything like that of the main sequence. At most, only a very small fraction of the quiescent class could be misclassified main sequence objects. Might they be misclassified passive objects? Many of their properties are closer to passive than to main sequence galaxies. However, their dependence on mass and environment are quite different (Figure 8),  as is their gas content.  In a sample of quiescent and passive galaxies within the ALFALFA survey volume, matched in redshift and mass distribution, HI gas is detected in 10.3\% of the quiescent galaxies but only 1.7\% of the passive objects. In similarly defined samples from the GASS survey, the fractions are 57\%  and 16\%. This is inconsistent with the quiescent objects being drawn from the passive population (but suggests, as one would expect, that some fraction of the objects in the passive class might be very low SSFR quiescent galaxies).  We conclude, therefore, that the great majority of the quiescent galaxies are correctly typed, and the results summarized above cannot be dismissed as being due to SFR measurement errors. That said, we do {\em not} claim that the quiescent and passive objects form separate populations. The dividing line between the two is set purely by the considerations described in Appendix A: below log(SSFR) = -12.2 we cannot reliably detect star formation in our sample galaxies. It is likely that the two groups form one continuous population differing only quantitatively in their SSFR values. Whether there is a distinct group of galaxies with absolutely no star formation is far beyond our ability to determine.
 
To understand the quiescent galaxies, we need a model of galaxy evolution with more content than the random assemblages of log normals used earlier. However, we do not need a full-blown numerical simulation; a toy model of galaxy formation based on the considerations mentioned in Section 1 will suffice.  As we pointed out earlier, galaxy evolution divides, roughly, into intrinsic and extrinsic effects. The intrinsic effects are very complicated, but they result in star--forming behavior which at recent times has been directly observed and characterized, i.e.  the Kennicutt-Schmidt Law. Extrinsic effects are not, in general, observable and- absent a full numerical model, are not calculable. We have no such model, and no completely successful model has yet been constructed.  However, if we ignore mergers, the result of these extrinsic processes is to deliver a net amount (infall - outflow) of {\em cold} gas to the star--forming regions of the galaxy with some unknown time dependence. Here we adapt the simplest possible such history: a constant rate of addition which begins at time $t_0 \ge 10^8$ years, and continues until a time $t_0 < t_1 \le 13.67 \times 10^9$ years.  We assume the the gas flows into the disk with  a radial density profile like that of the stars in disks today $\rho_{gas} \sim e^{-r/r_0}$.  From the data of Simard \etal (2011) we obtain mean values ${r_0} = 2.66\sqrt {{M_{gal}}/{{10}^{10}}} $, and a dispersion of about a factor of 2.0.  We take the parameters of the Kennicutt-Schmidt Law from Kennicutt (1998)- more recent compilations (e.g Daddi \etal 2010) are still consistent with these values- which can be expressed as $\Sigma _{SFR} = 1.5 \times {10^{ - 10}}\Sigma _{gas}^{1.4}$, where $\Sigma _{SFR}$ is the surface density of star formation in solar masses per year per square parsec, and  $\Sigma _{gas}$ is the surface density of HI gas in solar masses per square parsec.

This model cannot deal with bulge formation, but it need not do so. Whether bulges are formed very early in galactic history, or present day bulges are the result of the last major merger between previous disk galaxies, the history of the {\em present} disk likely only begins after bulge formation is complete. After that time the bulge contributes mass but neither gas nor star formation. Since the typical galaxy in our mass range has a bulge mass fraction of about 0.5, we account for the bulge by shifting both the gas--to--mass ratio and the specific star formation rate of our model galaxies down by a factor of two.
  
\begin{figure}[tbph!]
\figurenum{13}
\plotone{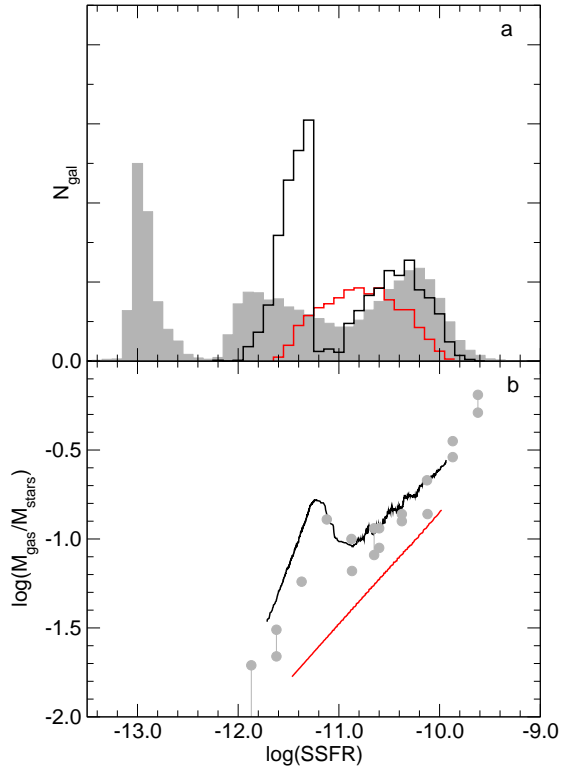}
\caption{Comparison of properties of galaxies in the mass range $10.0 \le log(Mass) < 11.0$ with the predictions of a simple model of galaxy evolution. a- distribution of SSFR; grey area- observations, red curve model evolution without disk stability cutoff in SFR, black curve- model evolution with disk stability cutoff. b- median gas--to--stellar mass ratio vs SSFR. grey points- observations, red points- model without disk stability cutoff, black points- model with disk stability cutoff.}
\end{figure}

 This model  is only a toy, but should not be a bad descriptor of at least the recent evolutionary history of the stellar populations of galaxies. Different infall histories or starbursts may change the short--term star--forming behavior, but will have little long--term effect.  In Figure 13 we compare the predictions of this model for SSFR distribution and gas content with the observations reported earlier, for galaxies in the mass range $10.0 \le logMass) < 11.0$.  The agreement is not good. The model reproduces, roughly, the SSFR distribution and gas content of the main sequence, but fails completely at lower values of SSFR, producing neither quiescent nor passive galaxies. However, our description of the star formation dependence on gas content is not yet complete. Kennicutt (1989), and   Kennicutt \& Martin (2001) have demonstrated that, below a critical density- probably related to the local Toomre Q disk stability threshold (Toomre 1964) - star formation rates fall precipitously. If the Q parameter is indeed the governing factor then, for a flat rotation curve, their results imply a critical density $\Sigma _{crit}({M_ \odot }p{c^{ - 2)}} = 0.69 \times 0.59 \times {V_{rot}}/R(kpc)$. From Bell \& de Jong (2001) we take a mean relation for disk galaxies $V_{rot} = 131 \times M_{10}^{0.23}$, where $M_{10}$ is the stellar mass in units of $10^{10}\Msun$, giving us the critical density as a function of galactic radius of $\Sigma _{crit}({M_ \odot }p{c^{ - 2}}) = 53.3M_{10}^{0.23}/R(kpc)$. The star formation rates below the critical density are not well determined. They are not zero: observations show that parts of galaxies below the critical density typically have small, isolated star forming regions. For lack of a better specification, we shall assume that below the critical density the SFR is 0.1 times the value predicted by the Kennicutt--Schmidt Law. 
 
 This more complete model is also shown in Figure 13, and it is is in much better  agreement with the observations. There is now a distinct main sequence, and a large group of quiescent galaxies with SSFR's about an order of magnitude lower than the main sequence. The gas to stellar mass ratio now flattens out at the location of the quiescent galaxies, with values of the gas to star formation ratio about a factor of 5 higher than on the main sequence. In fact, even the bump in gas--to--mass at $log(SSFR) \sim -11.1$ is reproduced. The reasons for this different behavior is explained in Figure 14, which presents the SFR histories of a sampling of galaxies of varying mass, $t_0$, and $t_1$ values, with and without stability quenching. Compared to models without a disk instability threshold, the star formation in models with a threshold drops more rapidly, as less and less of the galaxy is above the density threshold. Once all of the galaxy is below the threshold,  the global star formation efficiency stops falling, the gas consumption rate becomes very low, and the star formation rate drops very slowly. Galaxies move slowly in the Mass--SSFR plane, and pile up near the locus at which they made the transition. Although gas density in the disk drops with radius, so does the critical density, so the entire galaxy makes the transition over a relatively short timescale.  {\em And note that, again, a wide range of randomly selected star formation histories has produced a narrow main sequence.}
 
 The agreement of model and observations in Figure 13 is, in fact, quite remarkable, given how simple our model is. It must be stressed that no tweaking of parameters has been done; we have taken star formation behavior straight from Kennicutt (1998) and Martin \& Kennicutt (2001), and galaxy structure straight from Simard \etal (2011). This strongly suggests that no additional astrophysics, and no additional gas flows are necessary to move galaxies from the main sequence to the quiescent state. By itself, the rather small dispersion in gas--to--mass ratios seen at a given SSFR in Figure 10 implies that the mass of cold gas observed by ALFALFA and GASS is equal to the mass of star--forming gas; little has been missed and little is beyond the star--forming region. The agreement of model and observations further implies that this gas mass is neither increased by significant inflows not depleted by outflows, {\em at least during the main sequence to quiescent transition} (more on this soon.) The meaning of the quiescent population is now clear: {\em quiescent galaxies are those in which the density of  cold disk gas is below the gas disk instability threshold.}  All of the evidence presented in this paper is consistent, then,  with the hypothesis that the evolution of disk galaxies away from the main sequence is driven by the depletion of gas due to star formation. The rate of evolution, and the appearance of the galaxies is governed by the Schmidt-Kennicutt Law, as modified by an instability threshold.

\begin{figure}[tbph!]
\figurenum{14}
\plotone{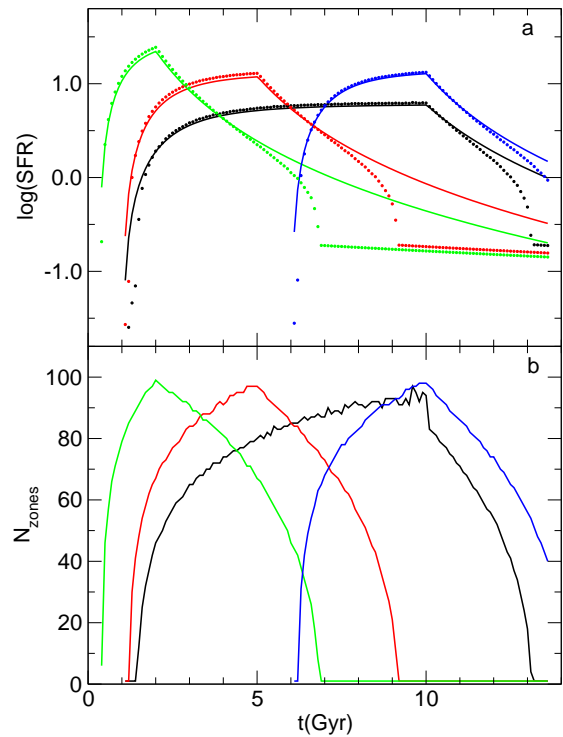}
\caption{Star formation histories of sample galaxies in our toy model. a- star formation vs time; solid lines are models without a disk instability threshold, circles are the same models with threshold. Parameters $t_0$,$t_1$ of the models are: green- $3\times10^8$,$3\times10^9$, red- $1\times10^9$,$5\times10^9$, black- $1\times10^9$, $1\times10^{10}$, blue- $6\times10^9$, $1\times10^{10}$. b- number of 100 pc radial zones with gas density above the instability threshold  vs time. Color coding is the same as in a.}
\end{figure}

 We have now solved the problem of the quiescent galaxies, but at the expense of raising another: if galaxy evolution due to star formation almost stops at, say, $logSSFR \sim -11.2$, how do galaxies make their way to lower values of SSFR, and how do they deplete their gas and enter the passive state? Two possibilities seem obvious: either (1) some galaxies, either contemporaneous with those entering the quiescent state or at an earlier epoch, somehow avoided the drop in star formation rate, and managed to turn their remaining gas into stars, or (2) another process is active which removes the remnant gas by other means. Possibility 1 seems unlikely, at least for disk galaxies. In a following paper (Oemler \etal in preparation) we will show that disk structure seems almost constant below $log(SSFR)<-11.0$, so that there is no reason to think that gas in the now--passive disks experienced very different conditions than those in quiescent disks today.  Ellipticals are, of course,  a different matter: It is well known that bulge--dominated galaxies are more prevalent among passive galaxies, and our toy model has no relevance to bulge formation. However, bulge-dominated galaxies comprise at most  30\% of the passive population, so different star formation astrophysics in bulges cannot be more than part of the answer.
 
 Possibility 2 seems more promising, and many mechanisms are known that can expel gas from galaxies. The strengths of many are proportional to the star formation rate, and are unlikely to be effective at the very low SFRs of the quiescent galaxies, but not all have this limitation. Some Type I supernovae occur well after the cessation of star formation, as do winds from AGB stars. AGN's may as well, although it is not clear how important AGN's are in galaxies with small bulges. Any such mechanism must be able to explain Figure 9, which plotted, over the entire observed range of mass and environment, the ratios of main--sequence to other galaxies, and the ratio of quiescent to passive galaxies. These quantities are independent of each other, but there is a quite tight correlation between the two, at almost all values of mass and density. Thus, if we try to invoke separate explanations for the transformation of main sequence galaxies into quiescent objects and for the transformation of quiescent into passive objects, we  must contrive a tight connection between the two processes. Fortunately, there is a natural connection. Comparison of the model and observed SSFR distributions in Figure 13 implies that the transformation process we are seeking operates on a fairly long timescale, longer than a few Gyrs but shorter than a few 10's of Gyrs. If it were as short as the former, all quiescent galaxies at $log(SSFR) \sim -11.2$ would have been quickly transformed into quenched objects. If it were as long as the latter, all would still be at $log(SSFR) \sim -11.2$ rather than spread all the way down to the passive region. In our model, the evolutionary rate and present state of a galaxy depends mostly on the epoch at which inflow ends. If certain environments and masses skew this epoch towards earlier values, more galaxies will have entered the quiescent state by now, {\em and, on average, they will have done so at an earlier time.} Thus, there will have been more time for the slowly working quiescent to passive transformation process to work, and more will have made the transition. ({\em vid.} Kelson \etal 2016, which come to similar conclusions from a different direction.)
 
 \subsection{The Role of Environment}
 
 It is our contention that the post--main sequence evolution of disk galaxies is driven by two intrinsic processes: the depletion of HI gas through star formation followed by the expulsion of the remaining gas, probably by the combined effects of stellar winds and supernovae. However, this intrinsic process only begins after an extrinsic event: the cessation of the net inflow of cold gas from outside. Now, it is possible, since it is the {\em net} inflow that matters,  that this is an intrinsic effect, due to an increase in outflows, perhaps driven by an AGN, rather than a cessation of inflows. However, unless such outflows were delicately balanced so as to only cancel the continuing inflows, rather than result in the rapid expulsion of all gas, such an event would lead to a much more rapid decline in star formation, and a much more rapid evolution of the galaxy towards the passive state, than is evidenced by the SSFR distribution in Figure 10. 
 
 Figure 8 confirms what has long been known: dense environments (and higher masses) push the galaxy population towards lower star formation rates. Given our model, these are, in fact, the trends one might expect. The environmental dependence on gas inflow was first proposed by LTC80, who pointed out that the observed star formation rates and gas content of spiral galaxies required continual infall of new gas if star formation was not to fade, and proposed that S0 galaxies were the product of the truncation of gas infall in dense environments. More recent models of a $\Lambda$ cold dark universe support such an idea. For example, Dekel \& Brinboim (2006) argued how the mass of the dark matter halo might govern the thermal properties of the infalling gas. Lower mass galaxies are fed by streams of cold gas, which can continue to feed star formation for extended epochs. In high mass galaxies the gas is shock heated to high temperatures at earlier times, after which it is more vulnerable to dispersal by winds generated by star formation and AGN's.  Feldmann \etal (2011, F11) have performed one of the first fine--scale numerical simulations of a large enough volume to simultaneously resolve the intrinsic processes of galaxies and also include a large enough volume to study the extrinsic processes driving evolution. They find that the evolutionary history of disk galaxies in a group are primarily driven by (a) mergers, which form the bulge component (b) suppression of gas infall within the group environment and (c) ram pressure stripping when galaxies pass through the group center. The stellar and gas content of the galaxies today depend on how long they have been in the group environment: those that entered first have evolved the most towards the passive state. 
 
 A significant role for ram pressure stripping of the disk gas is {\em not} consistent with our observations, since that would accelerate gas depletion and accelerate the transition of galaxies towards the passive state. However, the environment in the F11 simulation is denser than that experienced by the average galaxy. Although perhaps half of all galaxies are in groups, meaning that they have at least one neighbor of comparable or larger mass, the typical groups is significantly smaller than that studied by F11.  Cen (2014) has performed a larger--scale simulation of galaxies in their environments, containing 3000 galaxies in a considerably wider range of environments. He also finds that ram pressure stripping is important, but only for removing the large scale cold gas halos surrounding galaxies which resupply the disk. The subsequent evolution of the gas content of the disk is driven, as in LTC80, by star formation (see also  Rafieferantsoa \etal, 2015). We conclude, therefore, that ram pressure stripping of gas disks, as originally proposed by Gunn \& Gott (1972) will only be important for that small subset of galaxies living in the densest environments.

In our picture, the environmental dependence of galaxy stellar populations has two parallel components: the population of quiescent plus passive galaxies is driven by {\em how many} galaxies have had their outer gas halos removed sufficiently long ago for gas content to have fallen below the stability threshold, while the ratio of quiescent to passive galaxies is driven by  {\em how early} the halos were removed. It is reasonable to expect, and consistent with the models of Feldmann \etal and Cen, that the two trends should go together.

\section{Summary and Conclusions}

Beginning with rates inferred from hydrogen recombination lines, we have recalibrated star formation rates based on a combination of near--UV plus mid--IR flux, taking account of the fact that even old stellar populations have some flux in each of these bands. Using this, we examined the reliability of the largest available set of star formation determinations, that of SDSS galaxies by B04, as later modified and applied to the SDSS DR7 data release. We showed that the B04-DR7 values are generally quite good, but with systematic problems at low star formation rates. However, combining recalibrated B04-DR7 values with  rates derived  from the near--UV vs g band colors, we demonstrated that we can reliably measure specific star formation rates down to $log(SSFR) \sim -12$.

After dealing with selection effects, which bias optically selected galaxy samples against low SSFR objects, we have constructed a sample of about $10^5$ objects drawn from the B04-DR7 sample, which is biased in mass but unbiased in SSFR. This sample contains a very large number of galaxies with detected specific star formation rates well below that of the main sequence. The SSFR distribution of the sample suggests that the low SSFR objects might form a distinct set rather than the low tail of the main sequence, and we divide the sample into three subsets: {\em main sequence}, low SSFR objects, which we denote as {\em quiescent}, and {\em passive} objects with no detected star formation. We suspect that the passive and quiescent objects form a continuum, but since we are unable to resolve it we treat them as members of separate groups. 

We use measures of the galaxies' environments from Tempel \etal (2012), determinations of disk and bulge mass from Mendel \etal (2014), and gas content from the ALFALFA and GASS surveys, and we find the following trends among these populations:

\begin{itemize}

\item
averaged over all masses and environments, main sequence, quiescent, and passive galaxies each comprise about one third of the total galaxy population.

\item
Disk--dominated ($B/T<0.2$ galaxies are found only among main sequence galaxies, but few main sequence galaxies are bulge--dominated (($B/T>0.9$)

\item
The fractions of main sequence, quiescent, and passive galaxies are sensitive to environment and mass, main sequence fractions falling and quiescent and passive fractions rising at higher mass and density

\item
The mass and environment sensitivity of the three SSFR groups seem to depend on only one physical variable: the ratio of quiescent to passive galaxies is a tight function of the fraction of main sequence galaxies.

\item
There is a tight $\sigma \sim 0.3$dex correlation of SSFR with gas content; compared to main sequence galaxies, quiescent galaxies are overabundent in gas.

\end{itemize}

These trends place significant constraints on the histories of star formation in disk galaxies. We have shown, however, that one item- the existence of a tight correlation of mass with star formation rate, i.e. the main sequence- does {\em not} provide much constraint, and that any ensemble of star formation histories that satisfy two conditions- varying smoothly with time and decreasing at the present epoch- will display such a ``main sequence''. 

Recent work on the evolution of star formation has been dominated by one concept, that of the ``quenching'' of star formation. This is, perhaps, an unfortunate choice of terminology because ``quench'' is an ambiguous verb, which can be either transitive or intransitive, but is usually thought of as the former. Star formation quenches in a galaxy when star formation ceases, but that does not imply that it is quenched by an outside agency: the expulsion of gas by an AGN, the stripping of gas by ram pressure, or some other active force. The most basic reason why star formation ceases is that a galaxy runs out of gas, and the most basic reason why a galaxy runs out of gas is that it has turned all of it into stars. Unless provided with an endless new supply, that will happen to any galaxy, whether or not it is subjected to any additional gas removal process, as LTC80 pointed out long ago.

 Multiple pieces of evidence presented above suggest that the evolution of disk galaxies, in almost every environment, is driven by nothing more than gas exhaustion. The lack of disk--dominated quiescent and passive galaxies rules out any galaxy histories- stripping for example- in which the cessation of star formation is unconnected to the processes which built the galaxy's structure. Furthermore, the tight correlation of gas content and star formation rate, combined with the existence of a very large population of quiescent galaxies requires that the process behind gas removal acts very slowly, much slower than either ram pressure stripping or AGN-driven gas expulsion are likely to work. Since both of these processes work best when the gas density is low; they should, once started, drive the gas content to zero, rather than to the low but still significant levels seen in the quiescent galaxies.
 
 We have shown, using a simple but not unrealistic model for disk galaxies, that the properties of main sequence and quiescent galaxies are naturally explained by the observed dependence of star formation rate on density of gas in the disk. The evolution off the main sequence is governed by the Schmidt-Kennicutt Law, and the pile up of galaxies in the quiescent region is the result of the depression of star formation rates when disk gas density falls below a critical level, possibly set by the Toomre disk stability criterion. The only thing needed to start this evolution is for the infall of new gas into the disk cease, or at least drop to such a low level that it is unable to replace the gas that is locked up into new stars.
 
 In this picture, the evolution of disk galaxies is driven not by the (transitive) quenching of star formation due to some gas removal process, but rather by the (transitive) quenching of gas infall into disks, which precedes, by some billions of years, the (intransitive) quenching of star formation. The cutoff in gas infall provides a natural explanation for the environmental dependence of galaxy populations seen in Figure 8, since models show that such a cutoff tends to occur at earlier times in denser environments. The structure of disk galaxies present other challenges. If our model is correct, the correlation of bulge--to--disk ratio with SSFR, and therefore with disk evolutionary state implies a coherence between the bulge--building phase of galaxy evolution and the epoch at which gas infall ends. This may be so, but further consideration is beyond the scope of this paper.
  
 Finally, we should comment on the connection between these findings and the log--normal model for star formation histories which we have previously described (G13, Abramson \etal 2015, 2016). That model is characterized by star formation rates which peak at intermediate epochs and are characterized by two independent timescales. The star formation histories in Figure 14 share these characteristics, but not a precise log--normal shape. This is not surprising, for several reasons.  The observations which G13 fit were unable to detect the low SSFR levels of the quiescent galaxies and therefore do not include that phase in their evolution. Also, the shapes of the star formation histories which we have constructed in this paper  obviously depend on our choice of infall histories; the tophat form which we have used is unlikely to be an accurate description of actual infall. We are confident that we could more closely duplicate log normal histories by tuning the infall histories, but see little to be gained by such an effort. Whether the star formation histories of galaxies are accurately or only approximately log--normal in form remains an open question, but the existence of several timescales for galaxy evolution appears certain.
 
 \section{Acknowledgements}

B.V. acknowledges the support from an Australian Research Council Discovery Early Career Researcher Award (PD0028506)



\appendix

\section{Calibrating Galaxy Star Formation Rates and Struture}

\subsection{Star Formation Rates}

The largest published set of star formation rates for nearby galaxies comes from Brinchmann \etal (2004, B04), and is based on the $3\arcsec$ fiber spectroscopy from the Sloan Digitial Sky Survey ().  B04 first used the observed emission line strengths to divide the galaxies into 5 spectral categories based on line strength and the likelihood that the spectrum is contaminated by an AGN. The spectral energy distributions of objects with detectable emission lines and no AGN component were fit to a grid of stellar population models with a range of metallicity, ionization properties, and dust content, from which the star formation rate was derived. The SFR's of all other objects were derived from the strength of the 4000 Angstrom break, D4000,  by an empirical relation between break strength and SFR. Since the $3\arcsec$ fibers only sample a fraction of the typical galaxy, B04 apply empirically--derived aperture corrections based on the color distributions of the galaxies in SDSS imaging.

Salim \etal (2007) derived star formation rates for a very similar sample of SDSS galaxies using ultraviolet and optical photometry, and argued that the B04 SFR's of galaxies with no detected emission lines, which were based on the D4000 index, were systematically too high by several orders of magnitude. In response to this, Brinchmann and collaborators recalculated star formation rates, also basing them on the newer SDSS Data Release 7. We will use these values, which we denote as B04-DR7, in all of the following analysis.

Brinchmann \etal have probably done as careful and thorough a derivation of star formation rates as is possible from the SDSS data, but the reliability of the results is inevitably limited by the nature of that data: color distributions plus moderate S/N optical spectroscopy over a portion of the galaxy. As we shall see, the random errors in the B04-DR7 SFR's of vigorously star--forming galaxies are small, but the risk of systematic errors remain, particularly at the lower star formation rates which are our particular interest. To calibrate these possible errors we need independent, accurate star formation rates for a representative sample of the B04-DR7 objects. Since aperture corrections are a significant worry, the sample must, among other things, be representative of the redshift distribution of the B04-DR7 sample, most of which lie in the range $0.01 \le z \le 0.20$.

 All measures of star formation rates are indirect, but probably the best and least indirect are those based on the hydrogen recombination lines
 (Kennicutt 1998, K98 reviews all of the popular methods). Sufficiently accurate measures of at least two lines allow one to determine the extinction--corrected ionizing flux within HII regions, from which, with a few rather safe assumptions, the mass of young stars can be obtained. (Applying the method to entire galaxies is not free from error: if extinction with a galaxy is large and variable, the observed flux will come preferentially from the least--extincted regions and the average extinction correction will be underestimated.) The most accurate emission line measurements of a suitable galaxy sample are probably those of Moustakis \& Kennicutt (2006, MK06), which are derived from high signal--to--noise spectral scans over the entire face of a set of nearby galaxies.  We have derived \Ha luminosities from the  MK06  \Ha and \Hb fluxes after correcting for Galactic extinction using Schlafly \& Finkbeinder (2011).  Star formation rates are obtained using Equation 2 of K98, but adjusted for a Kroupa (2001) IMF: $SFR({M_ \odot }y{r^{ - 1}}) = 5.6 \times {10^{ - 42}}{L_{H\alpha }}(erg\,\;{\sec ^{ - 1}})$. 

These galaxies are not at all representative of  the B04-DR7 sample: they are all at quite low redshift and are all vigorously star--forming, however we can use them to calibrate another star formation measure which we can apply to a suitably large fraction of the B04-DR7 sample.  The most suitable such measure is that based on a combination of  ultraviolet and mid--infrared flux. Because they are photometric rather than spectroscopic measures, these fluxes encompass the entire bodies of galaxies rather than the smaller regions usually sampled spectroscopically. Also, because the UV is a measure of that fraction of the (reprocessed) ionizing hot star radiation which escapes the galaxies, while the mid--IR is a measure of the fraction that is absorbed by and warms the galactic dust, a suitable combination should be (relatively) immune to extinction within the galaxy.

There are several ways of combining these two fluxes; we follow the method applied by Kennicutt \etal (2009, K09), and simply  assume that $SFR_{UVIR} = C_{IR}L_{IR}+C_{UV}L_{UV}$,  where $L_{IR}$ and $L_{UV}$ are the luminosities in the UV and IR bands, solving for those values of $C_{IR}$ and $C_{UV}$ which minimize the scatter between $SFR_{UVIR}$ and $SFR_{Ha}$.

We use the subset of normal galaxies from MK06 which is complied in Table 2 of K09. For the infrared flux we take the IRAS (Rice \etal 1988, Soifer \etal 1989, Moshir \etal 1009 ) 25$\micron$ fluxes quoted by MK06. For the UV we take fluxes in the near--UV band from the GALEX Nearby Galaxy Atlas (Gil de Paz \etal 2007),  corrected as before for Galactic extinction. The scatter between \Ha and UV-IR derived star formation rates is minimized if we adapt
\begin{equation}
SFR_{UVIR} = 1.0\times10^{-43}(L_{25}+0.75L_{NUV})
\end{equation}
The agreement between \Ha and UV-IR star formation rates for the MK06 sample is presented in Figure A1a. The relation appears quite linear, with a scatter $\sigma(log(SFR)) = 0.12$.

To apply Equation A1 to the B04-DR7 sample we take \22m magnitudes from the WISE All Sky Data Release (Wright \etal 2010, Cutri \etal 2012), denoted W4. A comparison of WISE magnitudes and IRAS fluxes for objects observed by both shows that  $log(F_{25}) (Jy) = 0.4*(2.6-W4)$. For k corrections we use the 24$\micron$ k corrections from Rieke \etal (2009). Near--UV magnitudes are taken from the GALEX All Sky Imaging Survey Data Release 5 (Bianchi \etal 2011). We will also make use of the 2MASS (Skrutski \etal 2006) Ks magnitudes for our galaxies which are included in the WISE date release.  We correct K and NUV magnitudes for Galactic extinction using Schlafly \& Finkbeinder (2011) and apply k corrections from Chilingarian \& Zolotukhin (2012). 

\begin{figure}[tbph!]
\figurenum{A1}
\plotone{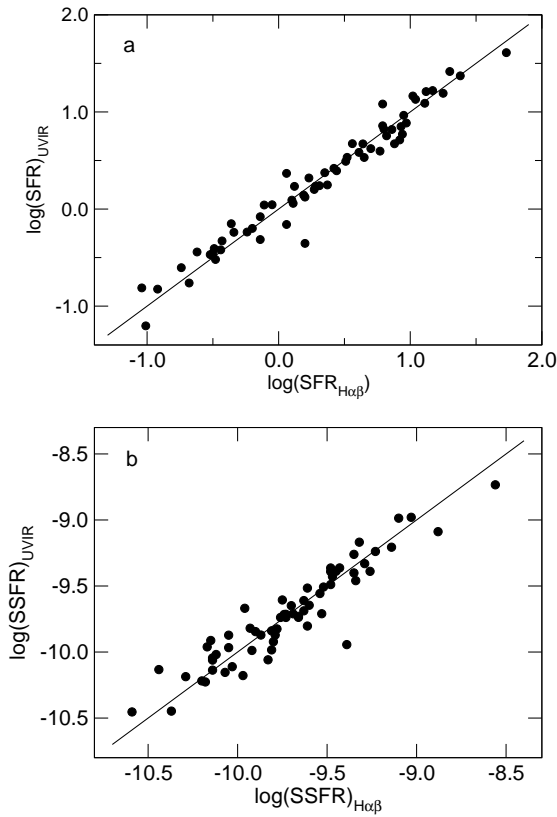}
\caption{a- Star formation rates of MK06 sample galaxies calculated following Equation A1, vs SFR from balmer lines. b- Specific star formation rates for the same sample.}
\end{figure}
Before proceeding with our recalibration of the B04-DR7 star formation rates, we must consider whether we can expect Equation A1 to be applicable at all levels of star formation, or more precisely, at all levels of {\em specific} star formation, since that is the quantity that we will be most interested in, and it is at very high or very low values of SSFR that Equation A1 is most likely to break down. Figure A1b presents the relation for the MK06 sample; it is quite linear over the range sampled, but that range only encompasses the ``main sequence''. There is reason to expect the relation to fail at low levels of UV and IR flux. As Schiminovich \etal (2007) have pointed out, old stellar populations as well as young populations contain hot stars- post--AGB stars for example- which will produce UV flux that is not indicative of star formation. The same is true in the mid--infrared, since most dust heating comes from the same hot stars  which produce the UV (Wang \& Heckman 1996). Given the uncertainties in the hot star populations of old stellar populations, it is safest to estimate this effect empirically.

\begin{figure}[tbph!]
\figurenum{A2}
\plotone{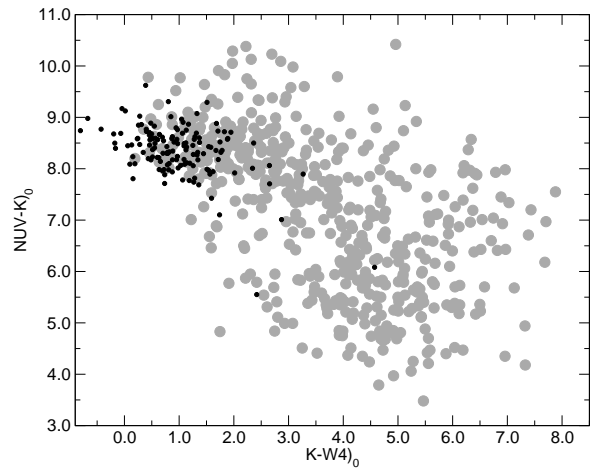}
\caption{Distribution of galaxies in the NUV-K vs K-W4 plane. Grey points- all nearby B04-DR7 objects with well--detected WISE W4 magnitudes, black points- ATLAS3D ellipticals with no detected molecular gas.}
\end{figure}

Figure A2 plots the distribution of two colors NUV-K and K-W4, for B04-DR7 galaxies with $z<0.015$ and well--detected W4. These two colors compare the UV and IR indicators of hot stars to the cool stellar component which dominates in the K band. There is a trend in the expected direction, but the distribution is very broad,  presumably due to the effect of internal extinction on the NUV flux. Also plotted, as black  points, is a set of  galaxies which are expected to have minimal star formation: ellipticals in which the ATLAS3D survey  (Young \etal 2011 ) detected no molecular gas. With a few exceptions- which may be true star--forming galaxies- these points cluster tightly around the region $NUV-K=8.5$, $K-W4=0.85$. We take these values to represent the UV and IR emission due to old stars, and therefore can produce UV and IR magnitudes corrected for this contamination as:
\begin{align}
\nonumber
NUV-K<8.5:&\\
NUV_{corr} &= -2.5log(10^{-0.4NUV}-10^{-0.4(K+8.5)})\\
\nonumber
 K-W4 > 0.85:\\
W4_{corr} &= -2.5log(10^{-0.4W4}-10^{-0.4(K-0.85)})	   
\end{align}
Beyond $NUV-K=8.5$ and $K-W4=0.85$ $NUV_{corr}$ and $W4_{corr}$ are set to arbitrarily large values. From these we calculate UV and IR luminosities to which we apply Equation A1 and we compare, in Figure A3, specific star formation rates so derived with those of B04-DR7. For reasons we will discuss shortly, we limit the sample to objects with disk inclination angles (from Simard \etal 2011) less that $60\degr$. The solid line has
\begin{align}
\nonumber
log(SSFR)_{UVIR} &= log(SSFR)_{B04corr}\\
                          &= 1.07log(SSFR)_{B04} +0.64
\end{align}
The scatter, $\sigma(log(SSFR))$, about this line ranges from 0.18 at $log(SSFR)=-10.0$ to 0.56 at $log(SSFR)=-11.5$. Thus, the B04-DR7 star formation rates are quite good for vigorously star--forming galaxies, but deteriorates significantly at lower specific star formation rates. This is hardly suprising: many of the latter had no detected emission lines, and star formation rates were estimated from D4000.

\begin{figure}[tbph!]
\figurenum{A3}
\plotone{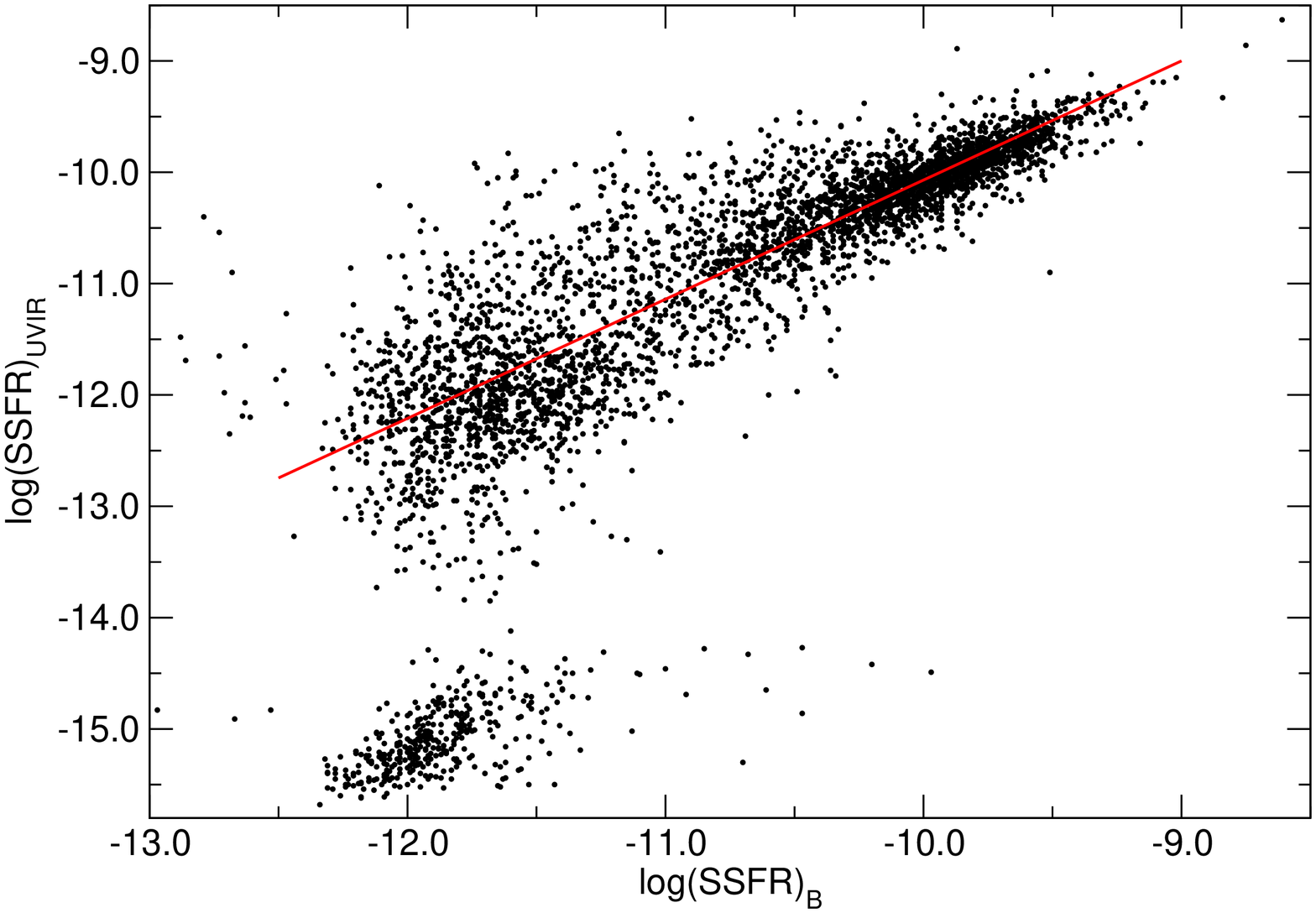}
\caption{Comparison of specific star formation rates from B04-DR7 with those derived from UV$+$IR photometry. The straight line is the best fit, as described by Equation 4.}
\end{figure}

This comparison with B04-DR7 is inadequate in one way: the depth of the WISE photometry in insufficient to span most of the redshift range of that sample. The median redshift of objects in Figure A3 is only 0.025, at which the 3\arcsec SDSS fibers only cover an area 3kpc in diameter. Thus, Figure A3  provides a stringent test of random errors in the B04DR7 extrapolations, but the limited redshift range is insensitive to systematic errors. To test for such errors we turn to the SWIRE survey (Londsdale \etal 2003), which used Spitzer to survey 6 small fields in the two galactic caps- three within the B04-DR7 area- to much greater depth. Intercomparison between the SWIRE 24$\micron$ photometry and the WISE and IRAS photometry show that $F_{25}(IRAS) = 1.5F_{24}(SWIRE)$. We then apply equations A1 - A4 to the data, and compare the $B04_{corr}$ and UV--IR specific star formation rates versus redshift in Figure A4. Galaxies with $log(SSFR)_{B04corr}>-11.0$, which have a smaller scatter, are shown as  filled circles, others as open circles. There is no apparent trend with redshift, implying that the B04-DR7 aperture corrections have no systematic errors.

\begin{figure}[tbph!]
\figurenum{A4}
\plotone{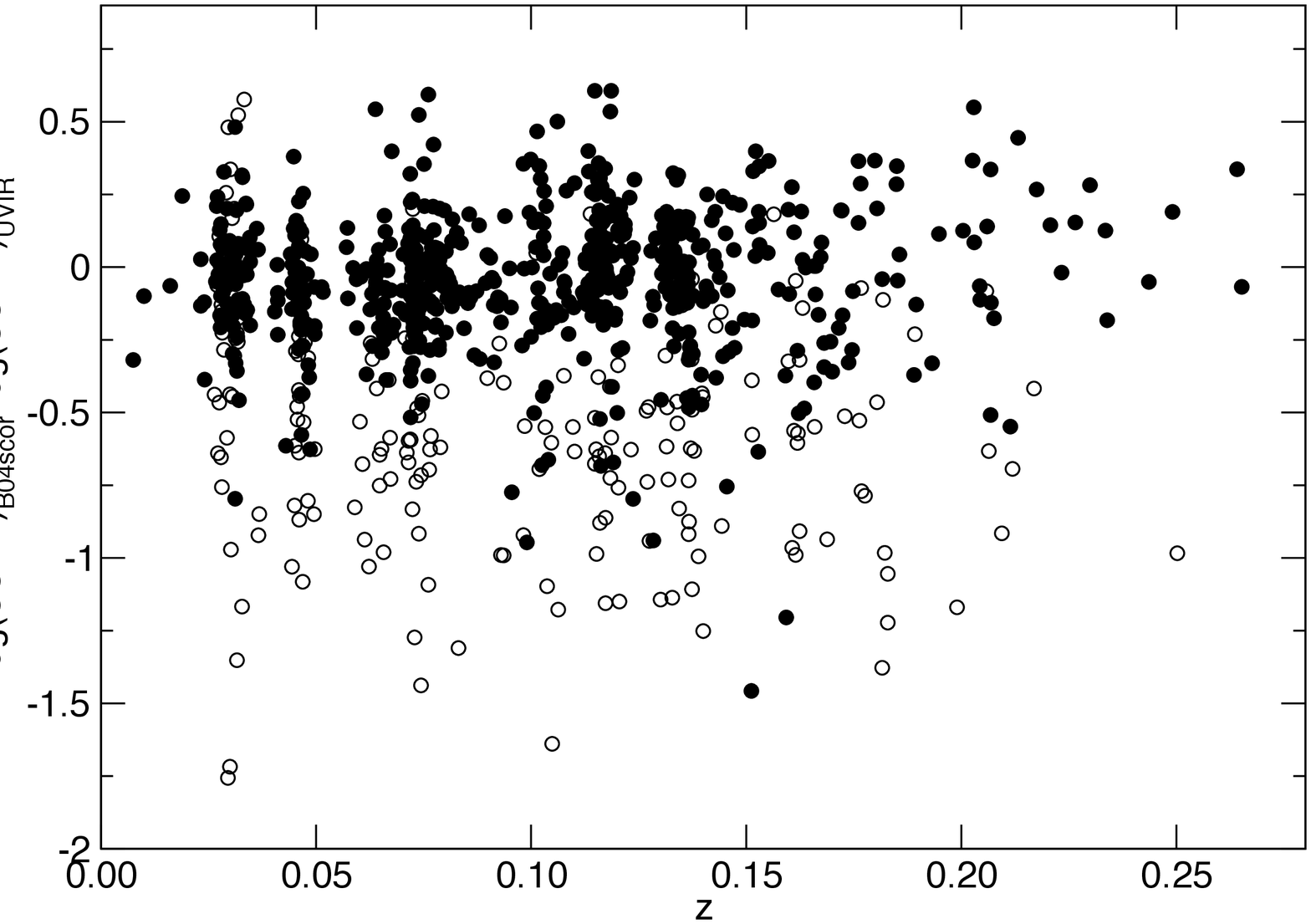}
\caption{Scatter of $log(SSFR)_{B04corr}-log(SSFR)_{UVIR}$ versus redshift. Filled circles are objects with $log(SSFR)_{B04corr}>-11.0$.}
\end{figure}

The above presentation confirms that, once a small correction is applied, the B04-DR7 specific star formation rates are remarkably accurate, and free of systematic errors. The SSFR values of vigorously star--forming galaxies are sufficiently accurate for our needs; however, random errors do increase significantly at lower levels of star formation. Ideally, we would like to keep $\sigma(log(SSFR))$ less than a few tenths at all detectable levels of star formation, but to achieve this we need another accurate and independent measure of star formation. $SSFR_{UVIR}$ would be ideal, but adequately deep mid--infrared photometry does not exist over most of the B04-DR7 sample volume. However, GALEX UV photometry does go deep enough, and we shall examine one SSFR measure, based on NUV-g colors. We take NUV magnitudes from GALEX, and g magnitudes from B04, correcting, as before, for Galactic extincton, k corrections, and the UV flux from old populations following Equation A2.  Figure A5 presents the relation between $log(SSFR)_{UVIR}$ and $NUV-g)_0$ for the same sample used in Figure A3. The solid line is a best fit for $NUV-g)_0<4.6$, and is of the form
\begin{align}
\nonumber
log(SSFR)_{UVg} = &-8.36-1.42(NUV\negmedspace -\negmedspace g)+0.3809(NUV\negmedspace -\negmedspace g)^2\\
                            &-0.05517(NUV\negmedspace -\negmedspace g)^3
\end{align}
Given the sensitivity of UV flux to galactic extinction, it might seem puzzling how tight this correlation is. The reason, we surmise, is that both galactic star formation rate and dust content are well--correlated with gas content. The trend of observed NUV-g color with SSFR is the sum of two factors: extinction  and intrinsic NUV-g color, both of which are well-correlated with SSFR.

\begin{figure}[tbph!]
\figurenum{A5}
\plotone{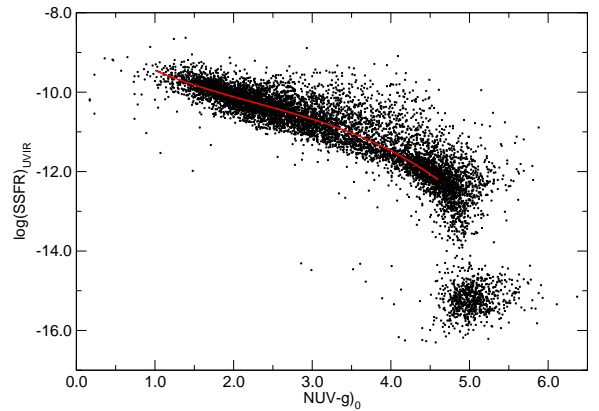}
\caption{$SSFR_{UVIR}$ versus NUV-g color for the B04-DR7 sample. The red curve is the adapted relation described by Equation A5.}
\end{figure}

We now define our SSFR measure
\begin{equation}
log (SSFR) \equiv 0.5*(\log {(SSFR)_{B04corr}} + \log {(SSFR)_{UVg}}
 \end{equation}
 and, in Figure A6, compare this with $SSFR_{UVIR}$. The two clouds of points to the left of $log(SSFR) = -12.25$ include objects for which $NUV-g)_0 \ge 4.6$; those in the lower cloud also have either $NUV-K \ge 8.5$ or $K-W4 \le 0.85$, or both. It is clear from this plot that we should assume that all values of $log(SSFR)$ less than -12.2 should be regarded as upper limits.

\begin{figure}[tbph!]
\figurenum{A6}
\plotone{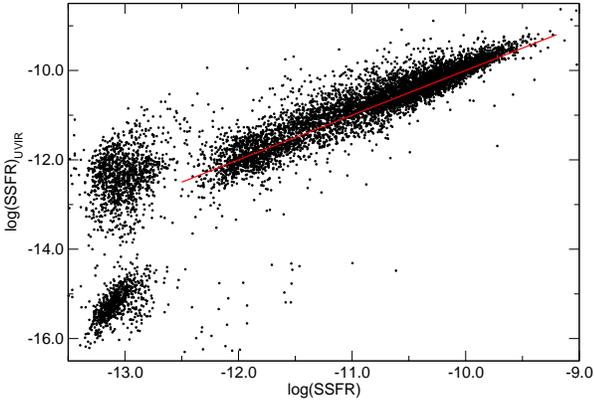}
\caption{Final SSFR values versus $SSFR_{UVIR}$.}
\end{figure}

Next, we must revisit the issue of internal galactic extinction. B04 attempt to correct for extinction as part of their analysis, we have argued above that our fit of $NUV-g)_0$ versus $SSFR_{UVIR}$ automatically accounts for extinction, and the IR+UV method is designed to be free of extinction effects. Thus, we would expect internal extinction to be unimportant, but Figure A7 shows that this is not true. We plot, for the galaxy sample used in previous figures, the median value of SSFR vs disk inclination angle for objects with $log(SSFR) \le -11.0$. It appears none of the methods properly account for extinction. B04-DR7 do the best job and NUV-g color does the worst, but even the IR+UV method has serious extinction problems. It is easy to understand why this is so: the sum of IR and UV flux may, {\em in the mean} account for all hot star ionizing flux, but only in the mean. The reradiation of absorbed UV flux by dust grains is a form of (energetic) scattering: flux is removed from the line of sight and reradiated isotropically. Thus, at high inclination angles more energy is removed from the line of sight in the UV than is returned to the line of sight in the IR, and Equation A1 will underestimate the total star formation rate. To correct for this, we fit the median of our final SSFR measure:
\begin{align}
\nonumber
\Delta (\log (SSFR) = &6.78 \times {10^{ - 3}}{I_{disk}} - 1.863 \times {10^{ - 4}}I_{disk}^2 \\
                                &+ 2.7543 \times {10^{ - 8}}I_{disk}^4
\end{align}
We cut our sample at $I_{disk}=85\fdg$ to minimize the uncertainty from very large corrections and apply this correction to our SSFR values. The final scatter in $log(SSFR)_{uvir}$ at fixed $log(SSFR)$ ranges from 0.13 at $log(SSFR)=-10.0$ to 0.16 at $log(SSFR) =-11.5$, quite satisfactorily small.

\begin{figure}[tbph!]
\figurenum{A7}
\plotone{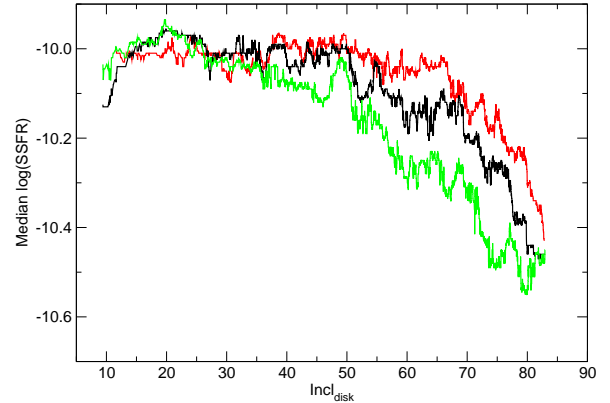}
\caption{Variation with disk inclination angle of the median SSFR of objects with $log(SSFR)  \le 011.0$. Red- corrected $B04_{corr}$ values, black- UVIR values, green UVg values.}
\end{figure}

 \end{document}